%% file: ms.tex
\documentclass[manuscript]{aastex}

\usepackage{amsmath}
\usepackage{amssymb}
\usepackage{graphics}

\def\TAME{\texttt{TAME} }
\def\lt{$<$ }
\def\gt{$>$ }
\def\le{$\le$ }
\def\ge{$\ge$ }
\def\teff{$T_\mathrm{eff}$}
\def\logg{log $g$}
\def\kms{$\mathrm{km~s^{-1}}$}
\newcommand{\FeI}{\ion{Fe}{1}}
\newcommand{\FeII}{\ion{Fe}{2}}

\newcommand{\ScII}{\ion{Sc}{2}}

\newcommand{\TiII}{\ion{Ti}{2}}
\newcommand{\MnI}{\ion{Mn}{1}}

\shorttitle{Abundances of Refractory Elements for G-type Stars with Extrasolar Planets}
\shortauthors{Kang et al.}

\begin{document}

\title{Abundances of Refractory Elements \\
       for G-type Stars with Extrasolar Planets}

\author{Wonseok Kang\footnote{\small Current Address : School of Space Research, Kyung Hee University, Gyeonggi-do 446-701, Republic of Korea, e-mail : wskang@khu.ac.kr}~ and Sang-Gak Lee} 
\affil{Astronomy Program, Department of Physics and Astronomy, Seoul National University, Seoul 151-742, Republic of Korea}

\and 

\author{Kang-Min Kim}
\affil{Korea Astronomy and Space Science Institute, Daejeon 305-348, Republic of Korea}

\begin{abstract}
We confirm the difference of chemical abundance between stars with and without exoplanet, as well as present the relation between chemical abundances and the physical properties of exoplanets such as planetary mass and semi-major axis of planetary orbit. 
We have obtained the spectra of 52 G-type stars with BOES (BOAO Echelle Spectrograph) and carried out the abundance analysis for 12 elements of Na, Mg, Al, Si, Ca, Sc, Ti, V, Cr, Mn, Co, and Ni. 
We first have found that the [Mn/Fe] ratios of planet-host stars are higher than those of comparisons in the whole metallicity range, and in metal-poor stars of [Fe/H] \lt -0.4, the abundance difference have been larger than in metal-rich samples, especially for the elements of Mg, Al, Sc, Ti, V, and Co. 
When examined the relation between planet properties and metallicities of planet-host stars, we have observed that planet-host stars with low-metallicity tend to bear several low-mass planets ($<~M_\mathrm{J}$) instead of a massive gas-giant planet. 

\end{abstract}

\keywords{stars: fundamental parameters --- stars: abundances --- stars: planetary systems}

\section{Introduction}

Since the first discovery of exoplanet around a normal star, \textit{51 Peg b} \citep{MQ95}, more than 500 exoplanets have been discovered. 
From the sufficient samples of planetary systems, the abundance studies for planet-host stars have been inspired to investigate the differences of elemental abundances between stars with and without exoplanets. 
At first, \citet{GO98} have suggested that planet-host stars tend to be metal-rich from spectral analysis of 8 planet-host stars, and the other studies confirmed that planet-host stars are metal-rich relative to normal field stars in the solar neighborhood \citep{GL01,SI01,SI03,BS03,LG03,SI04,SI05,BT06,EI07,SS08}.
The studies using a large number of samples revealed that the probability of finding a exoplanet was exponentially increasing with increasing metallicity \citep{FV05,JA10}. 
Despite the efforts to find the abundance differences for each element other than iron, however, most studies found no systematic difference of abundance for $\alpha$-elements between stars with and without exoplanet, except some studies \citep{BS03,GI06,RL06} that suggested the potential differences between two groups for some elements (e.g. Mg, Al, Si, Mn, V, Co, and Ni). 

In most cases of elemental abundance studies, however, the samples were simply divided into two groups of stars with and without exoplanets.
Most stars without known exoplanets have not been thoroughly examined for sufficient period. 
Even if the stars had been observed for long period, it would be still possible that those stars may have several planets whose masses are less than observational limit. 
Considering this incompleteness of samples with or without exoplanets, it should be inevitable that the reliability of statistical difference for chemical abundance between two groups of stars with planets and without known planets was reduced.
 
Nevertheless, it is worthwhile to investigate the relation between chemical abundance of host star and planet properties in the samples of planet hosts.
The number of low-mass exoplanets such as Neptunian planets are continuously increasing by more precise observation \citep{MS07,RL05,ME04,VB05}. 
The exoplanets in the wide range of their mass, therefore, makes it possible to examine the relation between planetary mass and chemical abundance of host star.
In this respect, \citet{SS08} have suggested that the detectability of Neptune-class planets may become higher in stars with low metallicity. 
It implies that the low-mass planets may be formed by a different process of planet formation from massive gas-giant planets, and suggests the probability that there is a relation between planetary mass and metallicity of host star.

On the other hand, in the core accretion scenario of planet formation, the amount of metals are important to the formation of not only terrestrial planets but gas-giant planets which require a lot of planetesimals for their core formation. 
The relation between metallicity of host stars and planet detectability have been presented in the abundance study of the uniform samples \citep{SI04,FV05,SI05,SS08,JA10} and supported by the theoretical study using core-accretion model \citep{IL04}. 
But there would be many other elements related with the process of planet formation. 
For example, the elements of Mg, Al, Si, Ca, and Ni are fairly abundant in the solar system relative to the amount of iron which represents the metallicity of planetary system. 
The Mg, Al, Si, and Ca are, especially, major elements for the condensation process in high temperature \citep{LO03} as well as show different trend of [X/H] from metallicity in low-metallicity stars via the galactic chemical evolution. 
Therefore, it is likely that these elements are related with the process of planet formation and it is useful to investigate the relation between planets and abundances of their host stars with low metallicity.

As mentioned above, stars without planets merely mean that they have no massive planets with short orbital period, because of the limit of detection method and observational time. 
Hence, we have focused on the samples of planet-host stars and properties of their planets that have been already well-confirmed, for example, hot Jupiters. 
In addition to this, if certain elements helped form planets in their formation process, their effect would be easily observed in the metal-poor stars because the circumstance of insufficient metals provided the poor conditions to form planets in core accretion scenario. 
And the chemical abundances of especially $\alpha$-elements such as Mg, Al, Si, and Ca, are more enhanced relative to the metallicity in low-metallicity stars. 
It can be, therefore, possible that detailed abundance analysis for low-metallicity stars gives a clue of condensation process in the planetesimal formation.

To achieve the goal in this study, we present the abundances of the 12 elements, Na, Mg, Al, Si, Ca, Sc, Ti, V, Cr, Mn, Co, and Ni for planet-host stars and stars without known exoplanets which are within 20 pc from the Sun, and the relation between the abundances of host stars and the properties of their plants. In Sect. 2, we introduce the samples and observations, and in Sect. 3 we present the abundance analysis method. In Sect. 4, we compare the abundances of host stars with the properties of planets around those host stars, and in Sect. 5, we discuss the difference of abundance results and evaluate the probability that the abundances of two groups are the same distribution using statistical test.

\section{The Data}

\subsection{Samples} 

The G-type stars among Planet-Host Stars (PHSs) were gathered from several exoplanet references \citep{BW06, EXOP}. 
Some controversial objects (HD 24040, HD 33636, and HD 137510) are excluded in the list of PHS and regarded as the comparison targets. 
We have constrained the PHS samples to the G-type stars with $\delta > -10^\circ$ and $V < 9.0$, in the conditions that can be observable and bright enough to obtain high S/N ratio spectra with 1.8 m telescope at BOAO (Bohyunsan Optical Astronomy Observatory). 
The comparison stars with no known planets were adopted from The Tycho-2 Spectral Type Catalog \citep{WM03}. 
These comparison stars also have $\delta > -10^\circ$ and $V < 9.0$ in the solar-neighborhood G-type stars within the distance of 20 pc from the Sun. 
For this abundance study, we have presented the results of 34 PHSs and 18 comparison stars in the list of G-type stars. 
\\

\subsection{Observations and Data Reduction}

The observations were carried out with the 1.8 m telescope at BOAO on 2008 and 2009. 
All spectra were obtained with BOES (BOAO Echelle Spectrograph) using 200 or 300 $\mu$m fiber.  
The observed spectra have a spectral resolution, R $\sim$ either about 30,000 (using 300$ \mu$m fiber) or 45,000 (using 200 $\mu$m fiber), and S/N ratios of higher than 150 at 6070\AA. 
The wavelength range of the spectra is from 3800\AA ~to 8800\AA, covering full optical region. 
The observational log and basic data of 52 targets are listed in \tablename~\ref{tbl:phslist}. 
In this table, the column 2 to 4 show observation date, exposure time (sec) and signal-to-noise ratio at 6070 \AA. 
The column 5 shows the radial velocity which was estimated by the difference between the observed and rest-frame wavelength of spectral lines in this study. 
And the column from 6 to 9 represent the stellar parameters determined by fine analysis which is explained in Sect. 3.
The column 10 indicates the spectral types of samples adopted from SIMBAD.

\begin{deluxetable}{lrrrrrrrrl}
\tabletypesize{\scriptsize}
\tablecaption{Target list\label{tbl:phslist}}
\tablewidth{0cm}
\tablehead{
\colhead{Star} &   
\colhead{Date}  &  
\colhead{Exposure}  & 
\colhead{S/N} &
\colhead{RV \tablenotemark{a}} & 
\colhead{$T_\mathrm{eff}$} & 
\colhead{log $g$} & 
\colhead{[Fe/H]} & 
\colhead{$\xi_\mathrm{t}$} &
\colhead{Type \tablenotemark{b} } \\
 &  & [sec] & & [\kms] & [K] & [dex] & [dex] & [\kms] & 
}
\startdata
\multicolumn{10}{l}{\textbf{        34 Planet-Host Stars}}   \\[3pt]
    HD 10697 & 2008-12-18 &   3600 & 400 &    -46.0$\pm$0.5 &  5662$\pm$80  &  4.07$\pm$0.11 &  +0.17$\pm$0.05 &  0.94$\pm$0.07 &           G5IV  \\ 
    HD 16141 & 2008-12-19 &   4800 & 450 &    -51.0$\pm$0.5 &  5755$\pm$83  &  4.17$\pm$0.11 &  +0.14$\pm$0.06 &  0.96$\pm$0.09 &           G5IV  \\ 
    HD 16400 & 2008-12-19 &   2400 & 500 &      9.1$\pm$0.5 &  4951$\pm$91  &  2.95$\pm$0.11 &  +0.12$\pm$0.05 &  1.34$\pm$0.09 &         G5III:  \\ 
    HD 17156 & 2009-02-06 &   4500 & 200 &     -3.2$\pm$0.4 &  6079$\pm$98  &  4.24$\pm$0.11 &  +0.21$\pm$0.06 &  1.06$\pm$0.10 &             G5  \\ 
   BD+20 518 & 2009-02-07 &   4800 & 250 &     -5.3$\pm$0.3 &  5540$\pm$83  &  4.31$\pm$0.10 &  +0.30$\pm$0.07 &  0.73$\pm$0.08 &             G5  \\ 
    HD 20367 & 2008-12-19 &   3600 & 500 &      6.5$\pm$0.3 &  6253$\pm$103 &  4.62$\pm$0.12 &  +0.16$\pm$0.06 &  1.25$\pm$0.08 &             G0  \\ 
    HD 28305 & 2008-12-18 &    360 & 450 &     38.4$\pm$0.3 &  4949$\pm$84  &  2.85$\pm$0.11 &  +0.21$\pm$0.08 &  1.42$\pm$0.09 &        G9.5III  \\ 
    HD 37124 & 2009-02-08 &   3600 & 250 &    -23.0$\pm$0.3 &  5551$\pm$79  &  4.48$\pm$0.11 &  -0.43$\pm$0.06 &  0.72$\pm$0.09 &         G4IV-V  \\ 
    HD 38529 & 2008-12-18 &   3600 & 350 &     30.4$\pm$0.2 &  5574$\pm$74  &  3.76$\pm$0.10 &  +0.32$\pm$0.09 &  1.26$\pm$0.10 &            G4V  \\ 
    HD 43691 & 2009-02-09 &   3600 & 200 &    -28.7$\pm$0.4 &  6229$\pm$101 &  4.25$\pm$0.11 &  +0.28$\pm$0.07 &  1.29$\pm$0.09 &             G0  \\ 
    HD 45350 & 2009-02-09 &   3600 & 200 &    -20.6$\pm$0.3 &  5636$\pm$84  &  4.37$\pm$0.10 &  +0.33$\pm$0.06 &  0.95$\pm$0.09 &             G5  \\ 
    HD 52265 & 2008-12-18 &   3600 & 450 &     54.1$\pm$0.3 &  6217$\pm$103 &  4.35$\pm$0.11 &  +0.24$\pm$0.06 &  1.16$\pm$0.11 &            G0V  \\ 
    HD 74156 & 2009-02-04 &   4500 & 300 &      3.7$\pm$0.4 &  6097$\pm$97  &  4.28$\pm$0.12 &  +0.13$\pm$0.06 &  1.10$\pm$0.11 &             G0  \\ 
    HD 75732 & 2008-04-27 &   1200 & 200 &     27.5$\pm$0.2 &  5246$\pm$80  &  4.26$\pm$0.09 &  +0.35$\pm$0.08 &  0.89$\pm$0.07 &            G8V  \\ 
    HD 75898 & 2009-02-09 &   4500 & 250 &     21.9$\pm$0.4 &  6063$\pm$93  &  4.19$\pm$0.11 &  +0.23$\pm$0.06 &  1.13$\pm$0.12 &             G0  \\ 
    HD 81040 & 2008-04-28 &   3600 & 250 &     49.2$\pm$0.3 &  5795$\pm$87  &  4.71$\pm$0.11 &  -0.04$\pm$0.07 &  0.80$\pm$0.08 &            G0V  \\ 
    HD 88133 & 2009-02-06 &   3600 & 250 &     -3.3$\pm$0.3 &  5414$\pm$80  &  4.03$\pm$0.10 &  +0.37$\pm$0.07 &  0.88$\pm$0.08 &           G5IV  \\ 
    HD 89307 & 2008-04-27 &   2700 & 200 &     23.0$\pm$0.3 &  6003$\pm$94  &  4.55$\pm$0.12 &  -0.11$\pm$0.06 &  1.08$\pm$0.11 &            G0V  \\ 
    HD 92788 & 2009-02-04 &   3600 & 350 &     -4.6$\pm$0.3 &  5751$\pm$88  &  4.38$\pm$0.10 &  +0.31$\pm$0.07 &  0.94$\pm$0.09 &             G5  \\ 
    HD 95128 & 2008-04-27 &    600 & 300 &     11.0$\pm$0.3 &  5900$\pm$89  &  4.36$\pm$0.11 &  +0.03$\pm$0.06 &  1.14$\pm$0.09 &            G1V  \\ 
   HD 104985 & 2008-04-24 &    900 & 250 &    -20.1$\pm$0.3 &  4623$\pm$70  &  2.27$\pm$0.12 &  -0.38$\pm$0.07 &  1.43$\pm$0.10 &          G9III  \\ 
   HD 106252 & 2008-04-28 &   3600 & 250 &     15.5$\pm$0.3 &  5976$\pm$92  &  4.56$\pm$0.12 &  -0.02$\pm$0.05 &  0.93$\pm$0.09 &             G0  \\ 
   HD 117176 & 2008-04-24 &    600 & 300 &      5.2$\pm$0.2 &  5520$\pm$74  &  4.00$\pm$0.11 &  -0.07$\pm$0.05 &  0.94$\pm$0.08 &            G5V  \\ 
   HD 143761 & 2008-04-27 &    600 & 300 &     17.7$\pm$0.3 &  5817$\pm$84  &  4.30$\pm$0.12 &  -0.21$\pm$0.06 &  0.91$\pm$0.08 &           G0Va  \\ 
   HD 149026 & 2009-02-04 &   4800 & 250 &    -17.9$\pm$0.5 &  6194$\pm$101 &  4.39$\pm$0.11 &  +0.32$\pm$0.07 &  1.16$\pm$0.10 &           G0IV  \\ 
   HD 154345 & 2008-04-24 &   1200 & 200 &    -46.6$\pm$0.3 &  5452$\pm$80  &  4.54$\pm$0.10 &  -0.08$\pm$0.05 &  0.46$\pm$0.10 &            G8V  \\ 
   HD 155358 & 2008-04-28 &   3000 & 250 &     -9.2$\pm$0.4 &  5944$\pm$91  &  4.27$\pm$0.13 &  -0.63$\pm$0.07 &  1.06$\pm$0.08 &             G0  \\ 
   HD 185269 & 2008-04-24 &   2400 & 250 &      1.3$\pm$0.3 &  6047$\pm$94  &  4.04$\pm$0.12 &  +0.14$\pm$0.07 &  1.28$\pm$0.11 &           G0IV  \\ 
   HD 186427 & 2008-04-24 &   1800 & 300 &    -27.4$\pm$0.3 &  5780$\pm$88  &  4.44$\pm$0.11 &  +0.09$\pm$0.06 &  0.95$\pm$0.09 &            G3V  \\ 
   HD 190228 & 2008-04-28 &   3000 & 200 &    -50.1$\pm$0.3 &  5278$\pm$68  &  3.78$\pm$0.11 &  -0.30$\pm$0.06 &  1.06$\pm$0.10 &           G5IV  \\ 
   HD 190360 & 2008-04-24 &   1200 & 250 &    -44.7$\pm$0.3 &  5533$\pm$81  &  4.31$\pm$0.10 &  +0.22$\pm$0.07 &  0.82$\pm$0.09 &       G6IV+...  \\ 
   HD 195019 & 2008-04-28 &   2400 & 350 &    -91.4$\pm$0.3 &  5820$\pm$86  &  4.19$\pm$0.11 &  +0.06$\pm$0.06 &  1.06$\pm$0.09 &         G3IV-V  \\ 
   HD 209458 & 2008-12-19 &   3600 & 300 &    -14.7$\pm$0.3 &  6127$\pm$101 &  4.43$\pm$0.12 &  +0.02$\pm$0.06 &  1.16$\pm$0.12 &            G0V  \\ 
   HD 217014 & 2008-12-18 &   1800 & 500 &    -33.1$\pm$0.2 &  5830$\pm$87  &  4.50$\pm$0.11 &  +0.24$\pm$0.05 &  1.05$\pm$0.09 &        G2.5IVa  \\[3pt] 
\multicolumn{10}{l}{\textbf{       18 Comparisons Stars} }   \\[3pt]
    HD 10307 & 2008-12-18 &   1800 & 450 &      4.4$\pm$0.2 &  5940$\pm$91  &  4.42$\pm$0.11 &  +0.05$\pm$0.05 &  1.14$\pm$0.09 &          G1.5V  \\ 
    HD 13974 & 2009-02-09 &    360 & 250 &      3.6$\pm$0.8 &  5944$\pm$92  &  4.43$\pm$0.12 &  -0.43$\pm$0.06 &  0.33$\pm$0.26 &          G0.5V  \\ 
    HD 24040 & 2009-02-08 &   3600 & 250 &     -9.3$\pm$0.3 &  5915$\pm$89  &  4.41$\pm$0.11 &  +0.24$\pm$0.10 &  0.98$\pm$0.09 &             G0  \\ 
    HD 26722 & 2008-12-18 &   1800 & 450 &     -8.2$\pm$0.4 &  5117$\pm$58  &  2.67$\pm$0.11 &  -0.10$\pm$0.08 &  1.43$\pm$0.10 &          G5III  \\ 
    HD 32923 & 2009-02-09 &    300 & 200 &     20.8$\pm$0.3 &  5702$\pm$81  &  4.18$\pm$0.11 &  -0.17$\pm$0.06 &  0.98$\pm$0.08 &            G4V  \\ 
    HD 33636 & 2008-12-19 &   4800 & 350 &      5.6$\pm$0.3 &  6040$\pm$93  &  4.61$\pm$0.12 &  -0.07$\pm$0.11 &  1.03$\pm$0.07 &        G0VH-03  \\ 
    HD 39587 & 2009-02-09 &    240 & 300 &    -12.1$\pm$0.7 &  5996$\pm$91  &  4.53$\pm$0.11 &  +0.00$\pm$0.04 &  1.00$\pm$0.07 &            G0V  \\ 
    HD 48682 & 2008-12-19 &   2400 & 500 &    -23.9$\pm$0.3 &  6144$\pm$96  &  4.43$\pm$0.12 &  +0.13$\pm$0.07 &  1.14$\pm$0.10 &            G0V  \\ 
    HD 50692 & 2008-04-25 &   1200 & 200 &    -15.0$\pm$0.4 &  5991$\pm$90  &  4.59$\pm$0.12 &  -0.12$\pm$0.05 &  1.04$\pm$0.09 &            G0V  \\ 
    HD 55575 & 2008-04-25 &   1080 & 250 &     84.8$\pm$0.3 &  5971$\pm$91  &  4.42$\pm$0.12 &  -0.25$\pm$0.04 &  0.97$\pm$0.09 &            G0V  \\ 
    HD 72905 & 2008-12-18 &   1500 & 450 &    -12.7$\pm$0.6 &  5920$\pm$88  &  4.57$\pm$0.11 &  -0.02$\pm$0.05 &  1.08$\pm$0.10 &         G1.5Vb  \\ 
    HD 84737 & 2008-12-19 &   1309 & 500 &      5.0$\pm$0.3 &  5958$\pm$89  &  4.19$\pm$0.11 &  +0.15$\pm$0.05 &  1.14$\pm$0.09 &         G0.5Va  \\ 
   HD 109358 & 2008-04-24 &    480 & 400 &      6.4$\pm$0.3 &  5912$\pm$88  &  4.48$\pm$0.12 &  -0.21$\pm$0.06 &  1.05$\pm$0.08 &            G0V  \\ 
   HD 110897 & 2008-04-27 &   1500 & 150 &     80.2$\pm$0.4 &  5805$\pm$86  &  4.32$\pm$0.12 &  -0.54$\pm$0.08 &  0.67$\pm$0.09 &            G0V  \\ 
   HD 137510 & 2008-04-27 &   2400 & 200 &     -6.8$\pm$0.4 &  5973$\pm$92  &  4.01$\pm$0.11 &  +0.29$\pm$0.12 &  1.24$\pm$0.08 &         G0IV-V  \\ 
   HD 141004 & 2008-04-24 &    480 & 400 &    -66.0$\pm$0.3 &  5951$\pm$91  &  4.28$\pm$0.12 &  +0.00$\pm$0.06 &  1.11$\pm$0.11 &            G0V  \\ 
   HD 161797 & 2008-04-27 &    120 & 250 &    -17.8$\pm$0.2 &  5621$\pm$78  &  4.06$\pm$0.10 &  +0.31$\pm$0.06 &  0.97$\pm$0.06 &           G5IV  \\ 
   HD 188512 & 2008-04-24 &    240 & 300 &    -39.5$\pm$0.2 &  5134$\pm$76  &  3.76$\pm$0.11 &  -0.10$\pm$0.07 &  0.79$\pm$0.09 &           G8IV  \\ 
\enddata
\tablenotetext{a}{Radial velocities which were obtained from center wavelengths via the measurements of equivalent widths}
\tablenotetext{b}{Spectral type from the SIMBAD database} 
\end{deluxetable}   
\clearpage 

\begin{figure*}[!ht]
\centering{
   \includegraphics[width=0.47\textwidth]{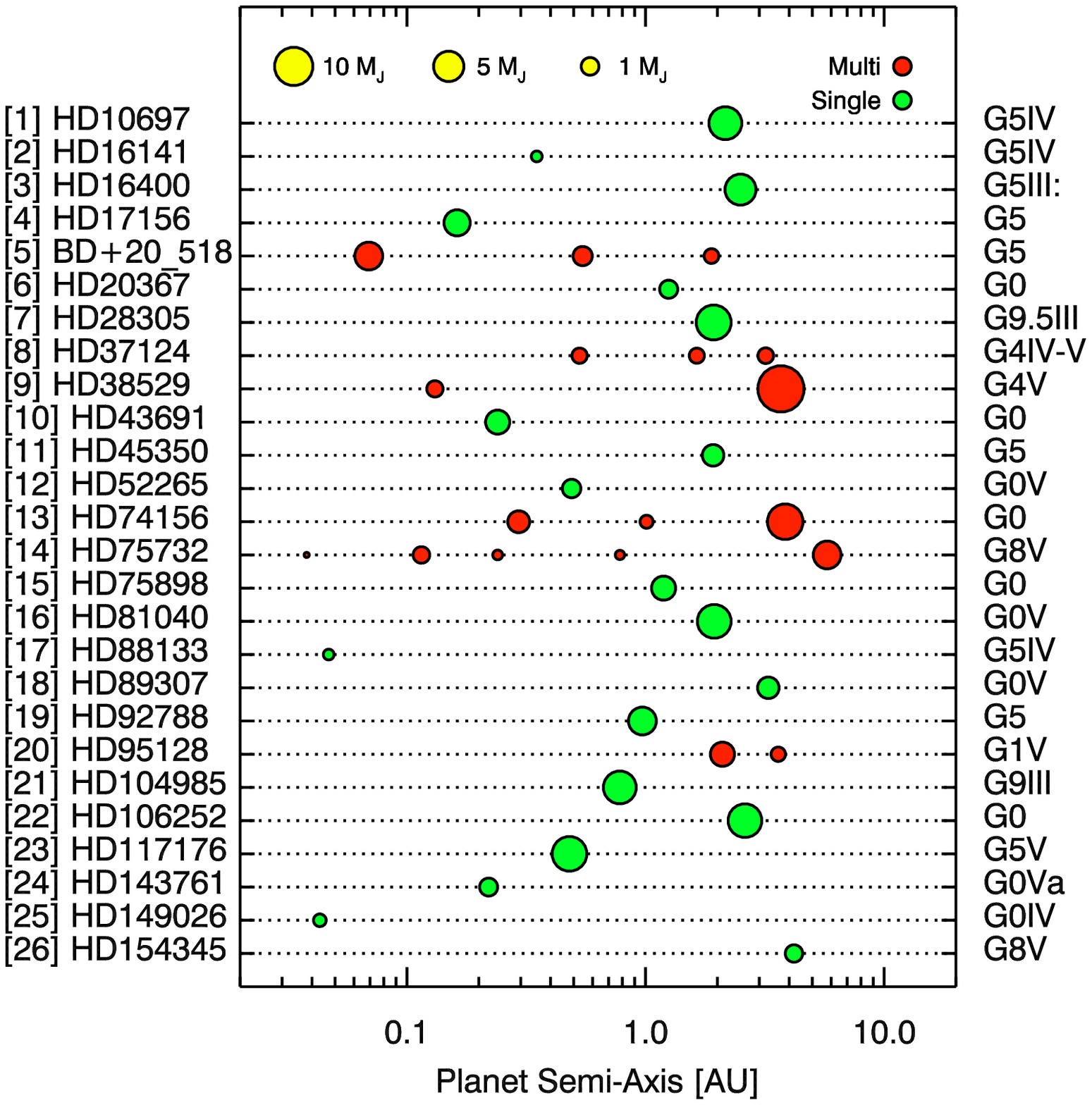} 
   \includegraphics[width=0.47\textwidth]{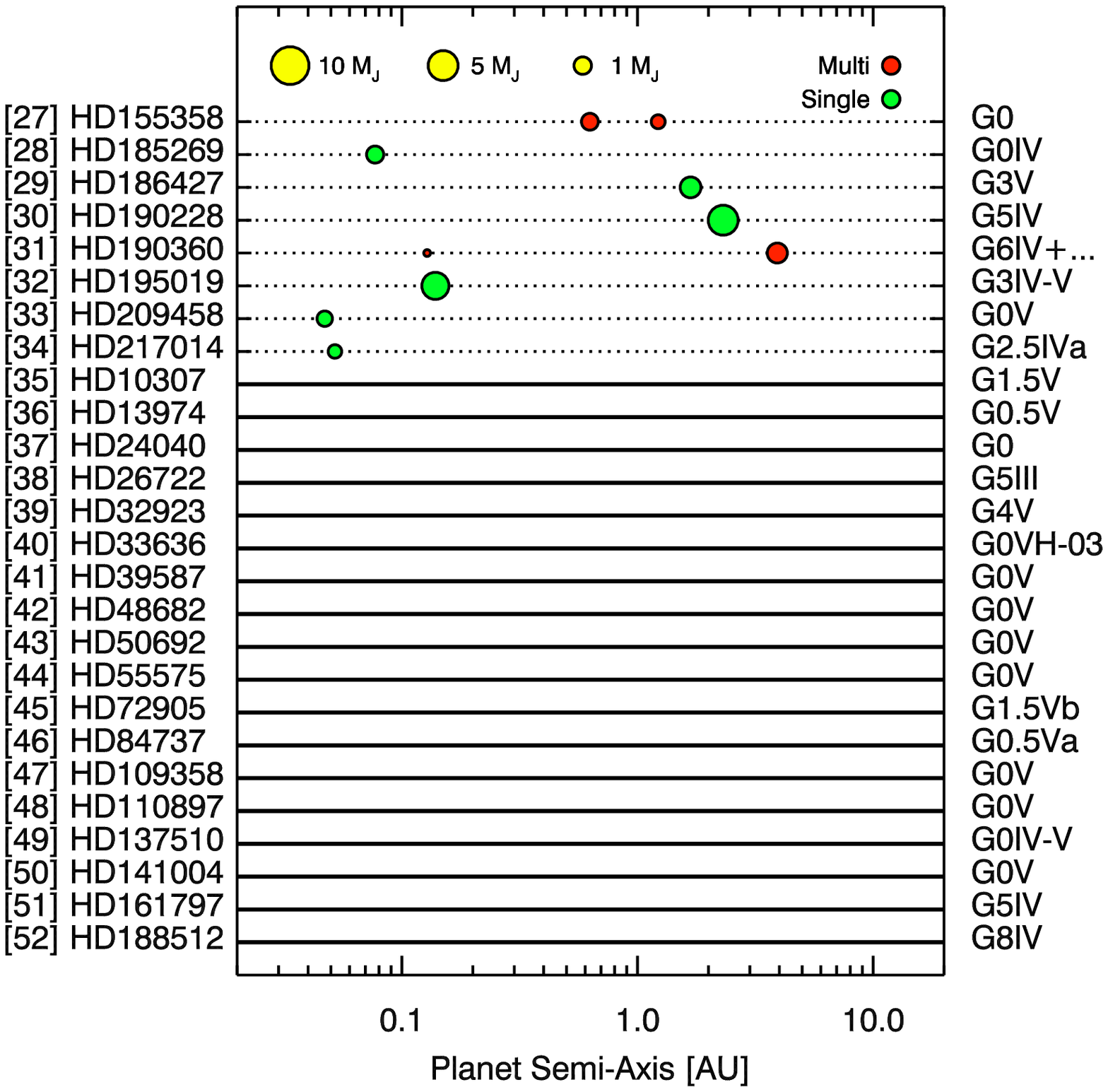} 
}
\caption{The sample list including the planet information. The size of circle represents planet mass, $(M_J \sin i)^{1/3}$ which means the relative size in the case that all planets had the same density. The red circles indicate the planets in multiple planetary system.}
\label{fig:targets}
\end{figure*}

Among observed 52 G-type stars, 34 stars have found planets around them by the radial-velocity method, and the semi-major axes and masses of their planets are shown in \figurename~\ref{fig:targets}, including 18 comparison samples. 
Since we are focusing on the relation between chemical abundances of host stars and properties of their planets, the observed samples of PHSs have outnumbered comparison stars. 
In \figurename~\ref{fig:targets}, the circle size represents the mass as $(M_\mathrm{J} \sin i)^{1/3}$, which means the relative size in the case that all planets had the same density, and the x-axis indicates the semi-major axis of a planet. There are 49 planets in these PHSs, and 10 planets are less massive than 10 $M_\mathrm{Nep}$ (the mass of Neptune is 0.054 $M_\mathrm{J}$) and only one planet of HD 190360 is as massive as 1 $M_\mathrm{Nep}$.

The reduction for observed spectra was carried out with IRAF \texttt{echelle} package to produce the spectra for each order of echelle spectrum. 
The echelle aperture tracing was performed using the master flat image which is combined with all the flat images. After aperture tracing, the flat, comparison, and object spectra was extracted from each image, with the same aperture reference of master flat image. In the flat-fielding process, the interference fringes and pixel-to-pixel variations of spectra were corrected. The wavelength calibration was performed with the ThAr lamp spectrum and the spectra of objects were normalized in each aperture using \texttt{continuum} task.

\section{Spectroscopic Analysis}

\subsection{Measurement of Equivalent Widths and Radial Velocity}

For the elemental abundances of stellar atmosphere, we use the method to measure the equivalent widths (EWs) of the relevant atomic lines. The measurement of EWs was carried out using the \TAME~(Tools for Automatic Measurement of Equivalent-widths) program, that we have developed for the abundance analysis with GUI interface. In order to run the \TAME, the spectra, the line list, and the parameter file are required. The \TAME~program is running in three steps: 

(1) the \TAME determines the local continuum of the spectrum in the wavelength range near the target line of the line list. The range to determine the local continuum is able to be adjusted. (2) the code finds the lines on the locally normalized spectrum using the derivatives of that. Using the second and third derivatives, the centers of absorption lines can be easily detected. (3) if there are blended lines near the target line, it fits those lines to the Gaussian/Voigt profile and measures the EW of the target line separated from nearby lines. 

For abundance analysis, the spectral lines of 12 elements (Na, Mg, Al, Si, Ca, Sc, Ti, V, Cr, Mn, Co, Ni) were referred from \citet{BT06} and \citet{GI06}. 
And in order to determine parameters of model atmosphere, the Fe lines are adopted from \citet{AG98} and VALD \citep{PK95,KP99}. 
For accurate estimation of abundances, these Fe lines have been examined using the solar spectrum obtained with BOES and we have selected the reliable lines that were not severely blended with nearby lines and were neither too strong nor too weak. 
And the value of oscillator strength (log $gf$ value) for each line was modified to the most recent value from VALD.  

\begin{figure}[!h]
\epsscale{0.7}
\plotone{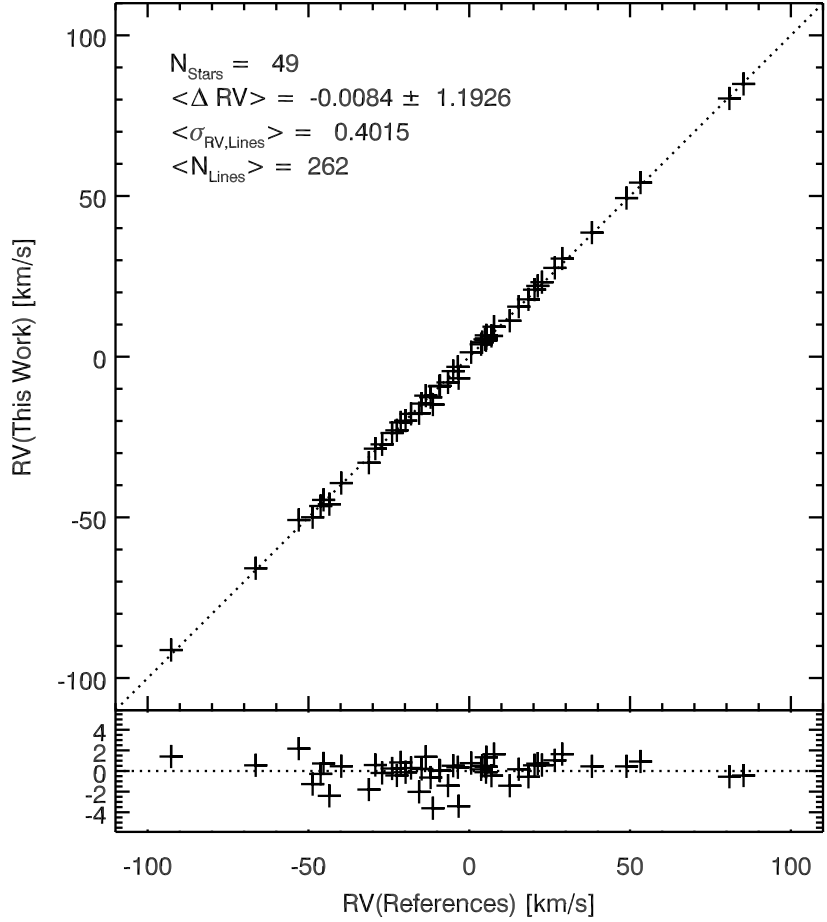}
\caption{The radial velocities of this work and reference (from SIMBAD) for 49 target stars.}
\label{fig:rvcomp}
\end{figure}

The \TAME~also estimates the center wavelength and FWHM of the target line. We calculated the radial velocities from the center wavelength of the fitting profile. 
For each star, the number of the lines which were used for the measurements is about 260 and the standard deviation within those lines is about 0.40 \kms.  
The radial velocities are listed in the column 5 of \tablename~\ref{tbl:phslist}. 
We compared these results with other studies for the radial velocity \citep{WI53,NM04,VF05} in \figurename~\ref{fig:rvcomp}, which shows a good agreement. 
The radial velocities of HD 17156 and BD+20 518 have been newly determined in this study, and HD 13974 is found to be a spectroscopic binary \citep{HM03}.

\subsection{Model Atmosphere}

\begin{figure}[!ht]
\epsscale{0.8}
\plotone{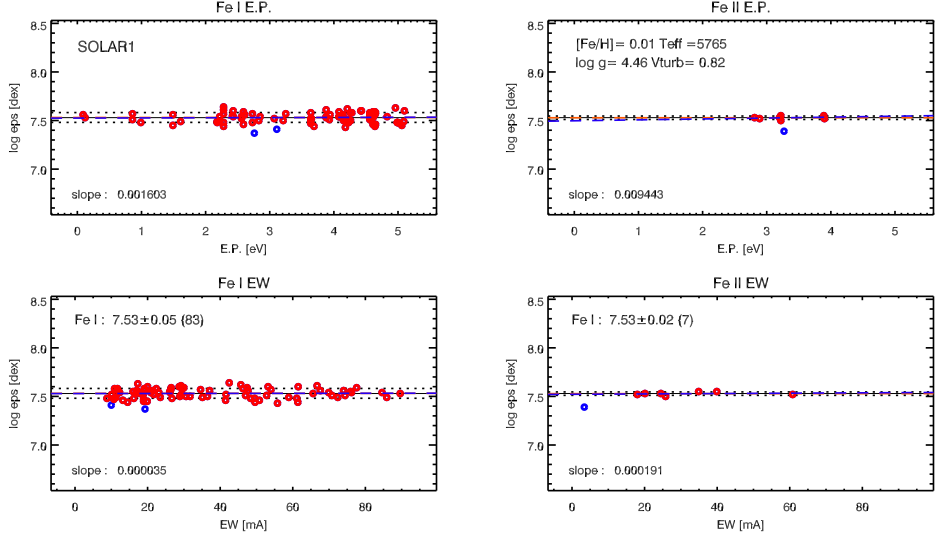}
\caption{The result of the fine analysis using Fe lines of the Sun. This plot shows the abundances for each Fe line (left:\FeI, right:\FeII). The top and bottom panel show the dependency of Fe abundance on excitation potential and EW of the line, respectively.} 
\label{fig:solarabu}
\end{figure}

We constructed the model atmosphere for each star using a model grid of Kurucz ATLAS9 without overshooting mode and with new opacity distribution functions (NEWODFs) \citep{ATLAS9, CK04}. 
The atmospheric parameters of Kurucz model atmosphere were determined by a self-consistent fine analysis, using the \FeI~and \FeII~abundances. 
The fine analysis was carried out by iterations to obtain the abundances of \FeI~and \FeII~lines on a model atmosphere to change their atmospheric parameters. 
The effective temperature and micro-turbulence parameters were adopted by iterating until the dependency of \FeI~abundance on excitation potential and EW was diminished.  
And the surface gravity was iteratively modified until the mean abundances of \FeI~and \FeII~to be the same. 
In order to reduce systematic errors from determination of model parameters, this process was automated by IDL code, and thus we could obtain model parameters for each star with the identical method. 
To test the process of this fine analysis, the atmospheric parameters of the Sun were determined using the solar spectrum obtained with BOES from the twilight on 10 May 2007, and the values were $T_\mathrm{eff}$ = 5765 $\pm$ 86 K, log \textit{g} = 4.46 $\pm$ 0.11 dex, [Fe/H] = 0.01 $\pm$ 0.05, $\xi_{t}$ = 0.82 $\pm$ 0.09 \kms. 
Using this atmosphere model, the solar Fe abundance of log $\epsilon$(\FeI ) = 7.53 $\pm$ 0.05 dex and log $\epsilon$(\FeII ) = 7.53 $\pm$ 0.02 dex was obtained as shown in \figurename~\ref{fig:solarabu}. 
This value has good agreement with the solar Fe abundance of log $\epsilon$(Fe) = 7.50 dex in \citet{GS98}. 

In this uniform way, the atmospheric parameters and Fe abundance for 52 stars were determined. 
We compared these atmospheric parameters determined by fine analysis with those of other studies \citep{CS01,VF05,GI06}. 
The differences of atmospheric parameters between this work and all references are about 48$\pm$70 K in \teff, 0.03$\pm$0.14 dex in log $g$, and 0.05$\pm$0.05 dex in [Fe/H]. 
As shown in \figurename~\ref{fig:cparam}, these parameters have a good agreement with those of three references. 


\subsection{Abundance Analysis and Uncertainties}

We determined the elemental abundances by LTE (local thermodynamic equilibrium) analysis relative to the solar abundances. 
After measuring the EWs of spectral lines for 13 elements including iron, the chemical abundances were derived with the 2002 version of MOOG code \citep{MOOG} using \texttt{abfind} driver. 
The Kurucz ATLAS9 model atmosphere determined by fine analysis of Fe I/II lines was used as the model atmosphere in MOOG code.
First, we derived the solar abundances of 12 elements from the EWs of elemental lines in the solar spectrum observed with the same spectrometer, BOES. 
And the elemental abundances for 52 stars were obtained with the same method to derive the solar abundance. 
Using the abundances of the Sun and 52 sample stars, we determined the differential abundances by comparing the abundances of 52 stars with those of the Sun. 

In this case, because the abundance was mainly determined by the measurement of EW, the systematic error could come from the measurement of EW. 
By comparing the line profile of spectrum with either Gaussian or Voigt profile, we excluded the EWs of lines that were severely blended with some unknown lines. 
And the local continuum level is another culprit for the error of EW measurement, so we also excluded the lines that were located in the crowded region which was difficult to determine local continuum in the spectrum. 

\begin{figure*}[!ht]
\centering{
   \includegraphics[width=0.32\textwidth]{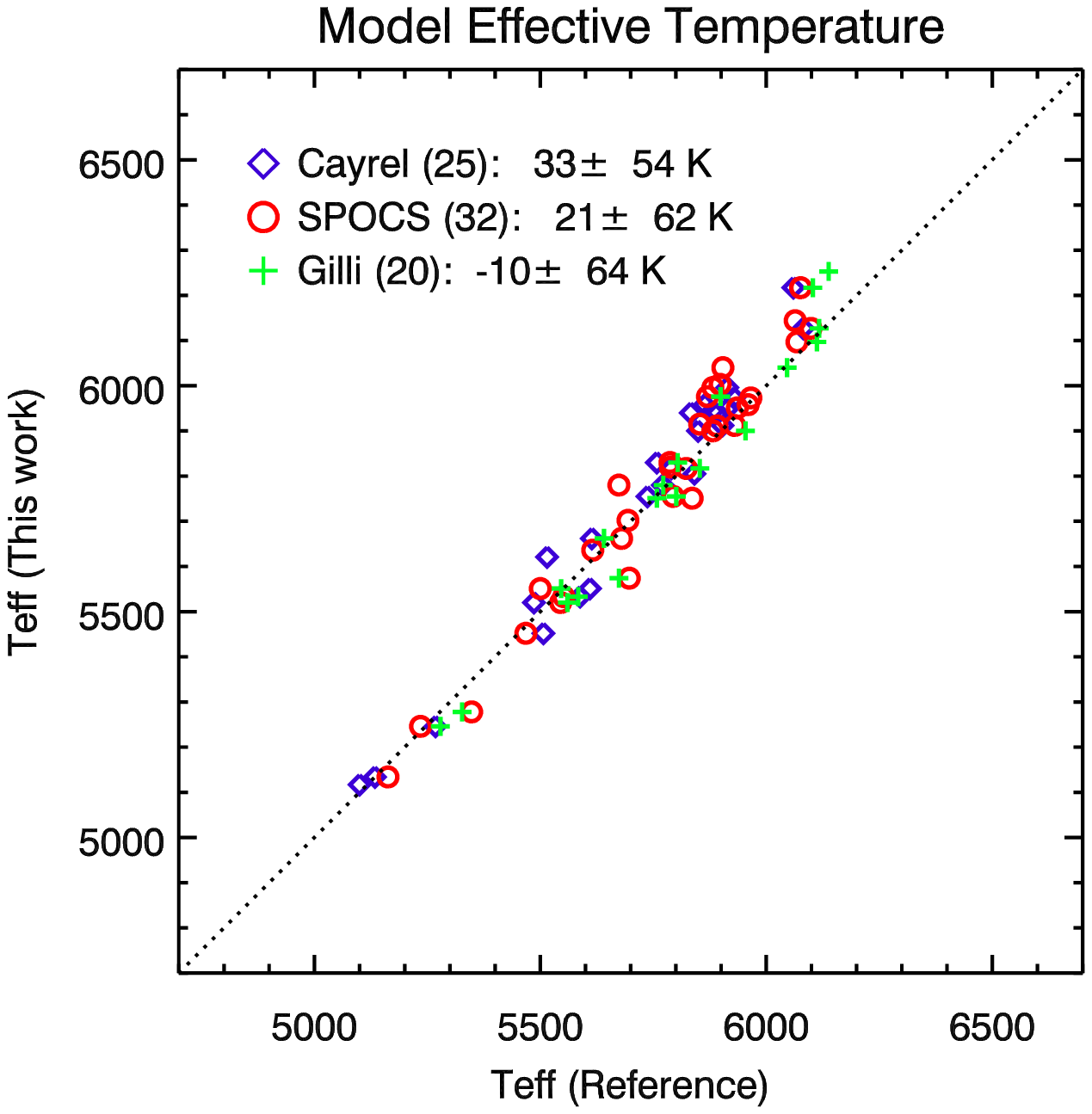} 
   \includegraphics[width=0.32\textwidth]{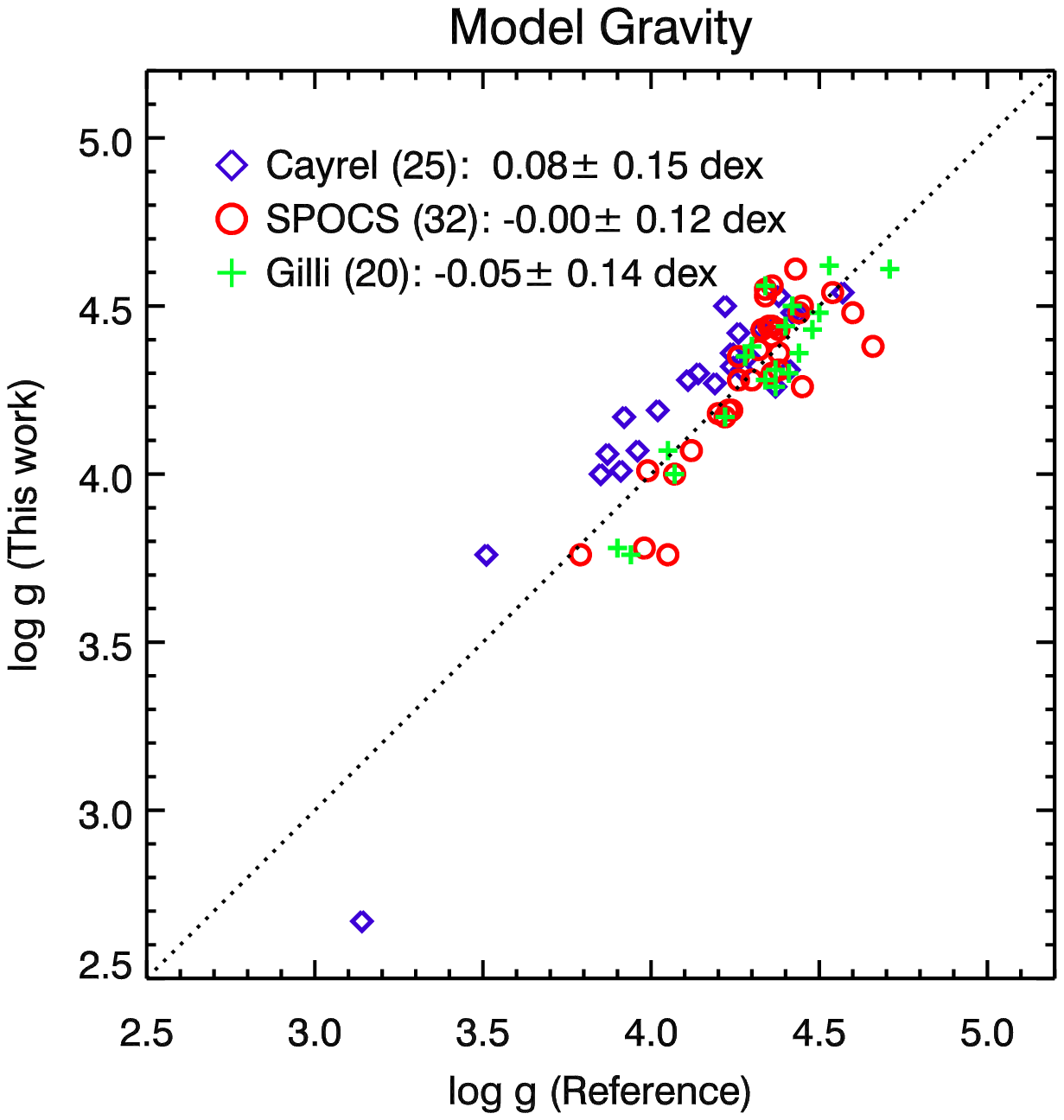} 
   \includegraphics[width=0.32\textwidth]{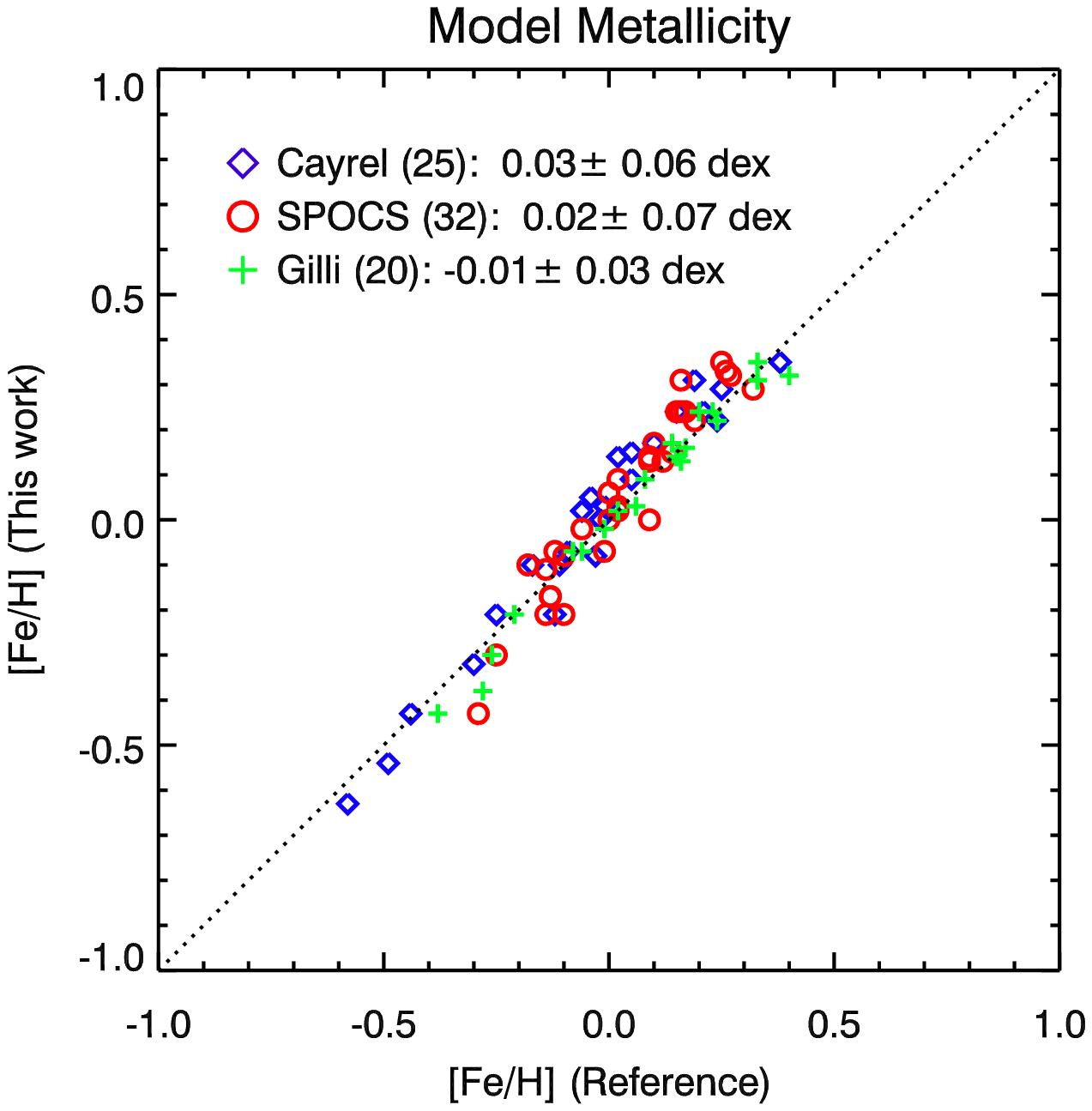}
}
\caption{The atmospheric parameters of this study and the three references \citep{CS01,VF05,GI06}. The numbers in  the parentheses indicate the number of the targets that were common in this study and in the references. The effective temperature, surface gravity, and metallicity parameters are determined in a good agreement with those of all references. The mean differences for effective temperature, surface gravity, and metallicity are less than 100 K, 0.1 dex, and 0.05 dex, respectively.  } 
\label{fig:cparam}
\end{figure*}

Systematic errors can also arise from the uncertainty of stellar parameters in use to make atmosphere model. 
If the parameters for model atmosphere were not correct, the unexpected errors of abundances could be occurred. 
Since this error was combined with EW measurement errors, it is hard to define separately but can be predicted through the variation of model parameters. 
Hence, we have examined the sensitivity of abundances on the model parameters. 
The stellar parameters of model atmosphere were changed by the amount of 100 K for \teff, 0.3 dex for [Fe/H], 0.3 dex for log $g$, and 0.3 \kms~for microturbulence($\xi_\mathrm{t}$). 
The abundance sensitivities for changing the parameters of model atmosphere are displayed in \tablename~\ref{tbl:modtest1}. 
The sample stars were selected in order that a testing parameter was gradually changed and simultaneously other parameters varied as small as possible. 
\tablename~\ref{tbl:modtest1} shows that most elements are insensitive to varying the stellar parameters within 0.13 dex. 
The abundances of \ScII~and \TiII~are the most sensitive to surface gravity because these elements are almost singly ionized around temperature of the solar atmosphere, to become these ionized lines are sensitive to gravity due to H$^-$ continuum opacity which is related with electron pressure. 
The largest variation of \ScII~and \TiII~abundance is, however, only about 0.13 dex with varying in 0.3 dex of log $g$. 
We estimated the uncertainties for effective temperature and micro-turbulence by probing the slope of Fe abundances for excitation potentials and EWs, and the uncertainty for surface gravity was adopted from exploring the abundance difference between \FeI\ and \FeII .
The uncertainties are listed in \tablename~\ref{tbl:phslist} with stellar parameters.
Considering that the uncertainty of \logg\ which is the most sensitive to abundances is within 0.15 dex (in \figurename~\ref{fig:cparam}), this abundance error which comes from the uncertainties for stellar parameters can be acceptable.

\clearpage

\begin{deluxetable}{lrrrrrrrrrrrrr}
\tabletypesize{\scriptsize}
\setlength{\tabcolsep}{0.02in}
\tablecaption{The sensitivity of abundance on the model parameters for each element}
\vspace{5pt}
\tablewidth{0cm}
\tablehead{
\colhead{Name} & \colhead{Na I} & \colhead{Mg I} & \colhead{Al I} & \colhead{Si I} & \colhead{Ca I} & 
\colhead{Sc II} & \colhead{Ti I} & \colhead{Ti II} & \colhead{V I} & \colhead{Cr I} & \colhead{Mn I} & 
\colhead{Co I} & \colhead{Ni I}  \\
~(\teff, [Fe/H], log $g$, $\xi_\mathrm{t}$) & & & & & & & & & & & & 
}
\startdata
\cutinhead{$\mathrm{T_{eff}}$ ; $\pm100$ K }

HD 209458                         &  -0.05 &  -0.05 &  -0.03 &  -0.03 &  -0.06 &  -0.01  &  -0.08 &  -0.01 &  -0.09 &  -0.05 &  -0.09 &  -0.07 &  -0.06 \\
~(6127, 0.02, 4.43, 1.16)         &  +0.04 &  +0.04 &  +0.03 &  +0.03 &  +0.06 &  +0.01  &  +0.08 &  +0.00 &  +0.09 &  +0.05 &  +0.09 &  +0.07 &  +0.06 \\[5pt]
 HD 106252                        &  -0.05 &  -0.04 &  -0.04 &  -0.02 &  -0.06 &  -0.00  &  -0.09 &  +0.00 &  -0.10 &  -0.06 &  -0.12 &  -0.06 &  -0.06 \\
~(5976, -0.02, 4.56, 0.93)        &  +0.05 &  +0.04 &  +0.04 &  +0.03 &  +0.06 &  +0.01  &  +0.09 &  +0.00 &  +0.09 &  +0.05 &  +0.11 &  +0.07 &  +0.06 \\[5pt]
 HD 154345                        &  -0.06 &  -0.03 &  -0.04 &  +0.00 &  -0.08 &  +0.00  &  -0.11 &  +0.01 &  -0.12 &  -0.07 &  -0.10 &  -0.06 &  -0.03 \\
~(5452, -0.08, 4.54, 0.46)  &  +0.06 &  +0.04 &  +0.05 &  +0.00 &  +0.07 &  -0.01  &  +0.11 &  -0.01 &  +0.12 &  +0.07 &  +0.10 &  +0.06 &  +0.04 \\[5pt]

\cutinhead{[Fe/H] ; $\pm0.30$ dex}

 HD 155358                        &  -0.02 &  -0.02 &  -0.01 &  -0.02 &  -0.03 &  -0.05  &  -0.03 &  -0.04 &  -0.04 &  -0.02 &  -0.04 &  -0.03 &  -0.03 \\
~(5944, -0.63, 4.27, 1.06)  &  -0.01 &  -0.01 &  -0.01 &  -0.00 &  -0.00 &  +0.06  &  -0.03 &  +0.06 &  -0.03 &  -0.02 &  -0.03 &  -0.02 &  -0.01 \\[5pt]
 HD 50692                         &  -0.04 &  -0.03 &  -0.02 &  -0.04 &  -0.04 &  -0.07  &  -0.03 &  -0.06 &  -0.03 &  -0.03 &  -0.05 &  -0.04 &  -0.03 \\
~(5911, -0.17, 4.44, 0.98)  &  +0.01 &  +0.01 &  -0.00 &  +0.03 &  +0.02 &  +0.08  &  -0.01 &  +0.08 &  -0.01 &  +0.00 &  +0.00 &  +0.00 &  +0.01 \\[5pt]
 HD 24040                         &  -0.03 &  -0.02 &  -0.00 &  -0.04 &  -0.04 &  -0.10  &  +0.00 &  -0.09 &  +0.01 &  -0.00 &  -0.01 &  -0.01 &  -0.02 \\
~(5915, 0.24, 4.41, 0.98)   &  +0.04 &  +0.03 &  +0.01 &  +0.05 &  +0.04 &  +0.10  &  +0.01 &  +0.09 &  +0.01 &  +0.02 &  +0.03 &  +0.03 &  +0.03 \\[5pt]

\cutinhead{ log $g$ ; $\pm0.30$ dex}

 HD 26722                         &  +0.04 &  +0.01 &  +0.01 &  -0.02 &  +0.04 &  -0.13 &  +0.01 &  -0.13 &  +0.01 &  +0.01 &  +0.01 &  -0.02 &  -0.02 \\
~(5117, -0.10, 2.67, 1.43)  &  -0.04 &  -0.02 &  -0.02 &  +0.02 &  -0.05 &  +0.12 &  -0.01 &  +0.12 &  -0.01 &  -0.01 &  -0.01 &  +0.02 &  +0.02 \\[5pt]
 HD 188512                        &  +0.06 &  +0.03 &  +0.02 &  -0.02 &  +0.09 &  -0.12 &  +0.02 &  -0.11 &  +0.01 &  +0.02 &  +0.02 &  -0.03 &  -0.02 \\
~(5134, -0.10, 3.76, 0.79)  &  -0.06 &  -0.04 &  -0.03 &  +0.02 &  -0.09 &  +0.11 &  -0.02 &  +0.10 &  -0.01 &  -0.02 &  -0.02 &  +0.04 &  +0.02 \\[5pt]
 HD 154345                        &  +0.06 &  +0.04 &  +0.03 &  -0.00 &  +0.11 &  -0.12 &  +0.02 &  -0.11 &  +0.01 &  +0.02 &  +0.01 &  -0.03 &  -0.00 \\
~(5452, -0.08, 4.54, 0.46)  &  -0.06 &  -0.04 &  -0.03 &  +0.01 &  -0.11 &  +0.11 &  -0.03 &  +0.10 &  -0.01 &  -0.03 &  -0.02 &  +0.03 &  +0.01 \\[5pt]

\cutinhead{$\xi_\mathrm{t}$ ; $\pm0.30$ dex}

 HD 137510                        &  +0.03 &  +0.03 &  +0.02 &  +0.03 &  +0.07 &  +0.09 &  +0.05 &  +0.11 &  +0.02 &  +0.03 &  +0.01 &  +0.02 &  +0.06 \\
~(5973, 0.29, 4.01, 1.24)   &  -0.03 &  -0.05 &  -0.02 &  -0.04 &  -0.08 &  -0.09 &  -0.05 &  -0.13 &  -0.02 &  -0.03 &  -0.01 &  -0.02 &  -0.06 \\[5pt]
 HD 161797A                       &  +0.02 &  +0.03 &  +0.02 &  +0.03 &  +0.06 &  +0.09 &  +0.05 &  +0.09 &  +0.06 &  +0.05 &  +0.11 &  +0.04 &  +0.07 \\
~(5621, 0.31, 4.06, 0.97)   &  -0.03 &  -0.03 &  -0.03 &  -0.04 &  -0.08 &  -0.10 &  -0.05 &  -0.11 &  -0.06 &  -0.05 &  -0.13 &  -0.04 &  -0.08 \\[5pt]
 BD+20 518                        &  +0.02 &  +0.02 &  +0.01 &  +0.02 &  +0.04 &  +0.07 &  +0.05 &  +0.06 &  +0.06 &  +0.04 &  +0.07 &  +0.03 &  +0.06 \\
~(5540, 0.30, 4.31, 0.73)   &  -0.02 &  -0.02 &  -0.02 &  -0.02 &  -0.05 &  -0.06 &  -0.06 &  -0.07 &  -0.06 &  -0.04 &  -0.10 &  -0.03 &  -0.06 \\
\enddata
\label{tbl:modtest1}
\end{deluxetable}

\clearpage

\section{Results}

\subsection{The Metallicities of Planet-Host Stars}

\begin{figure*}[!ht]
\epsscale{1}
\plottwo{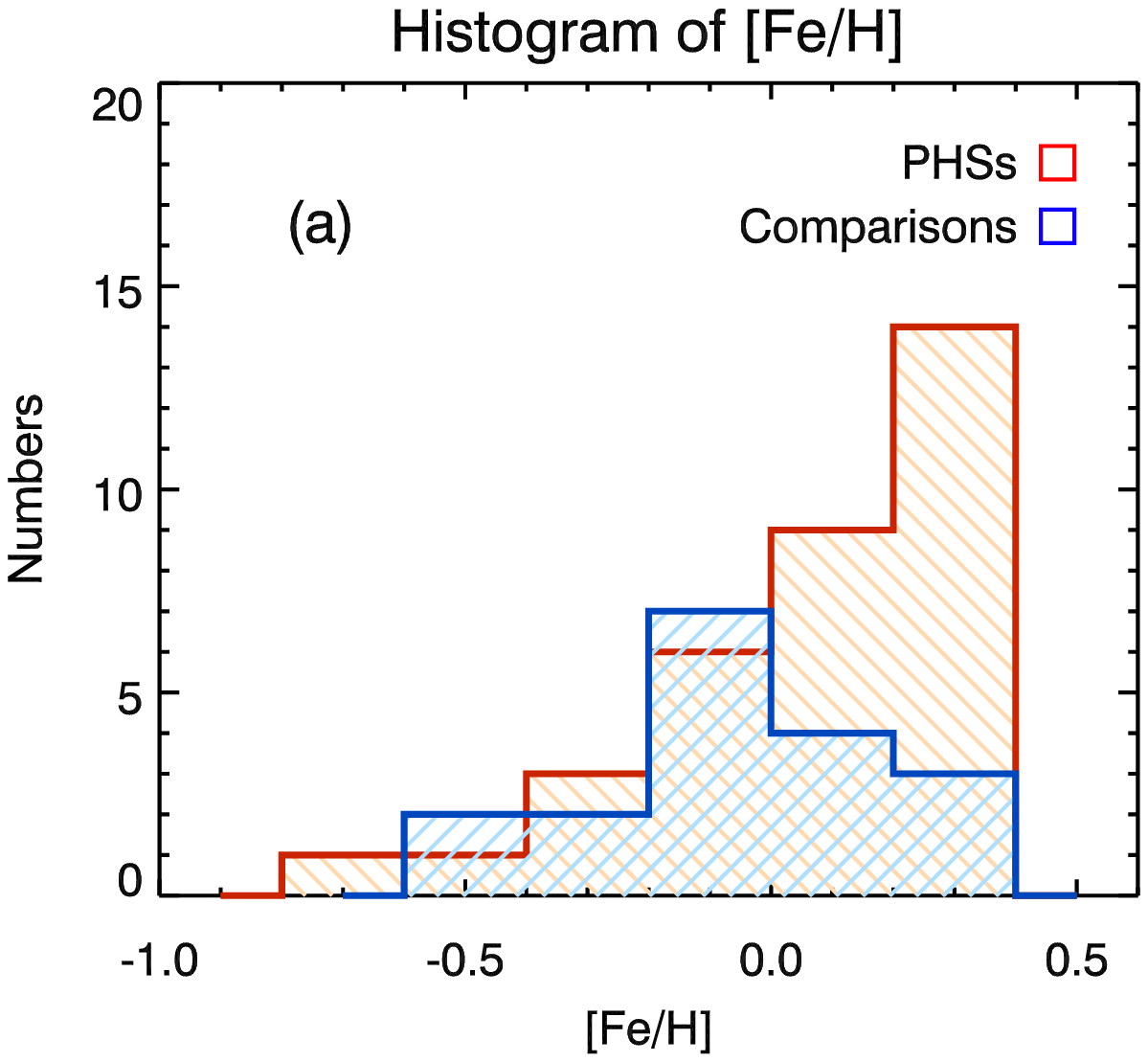}{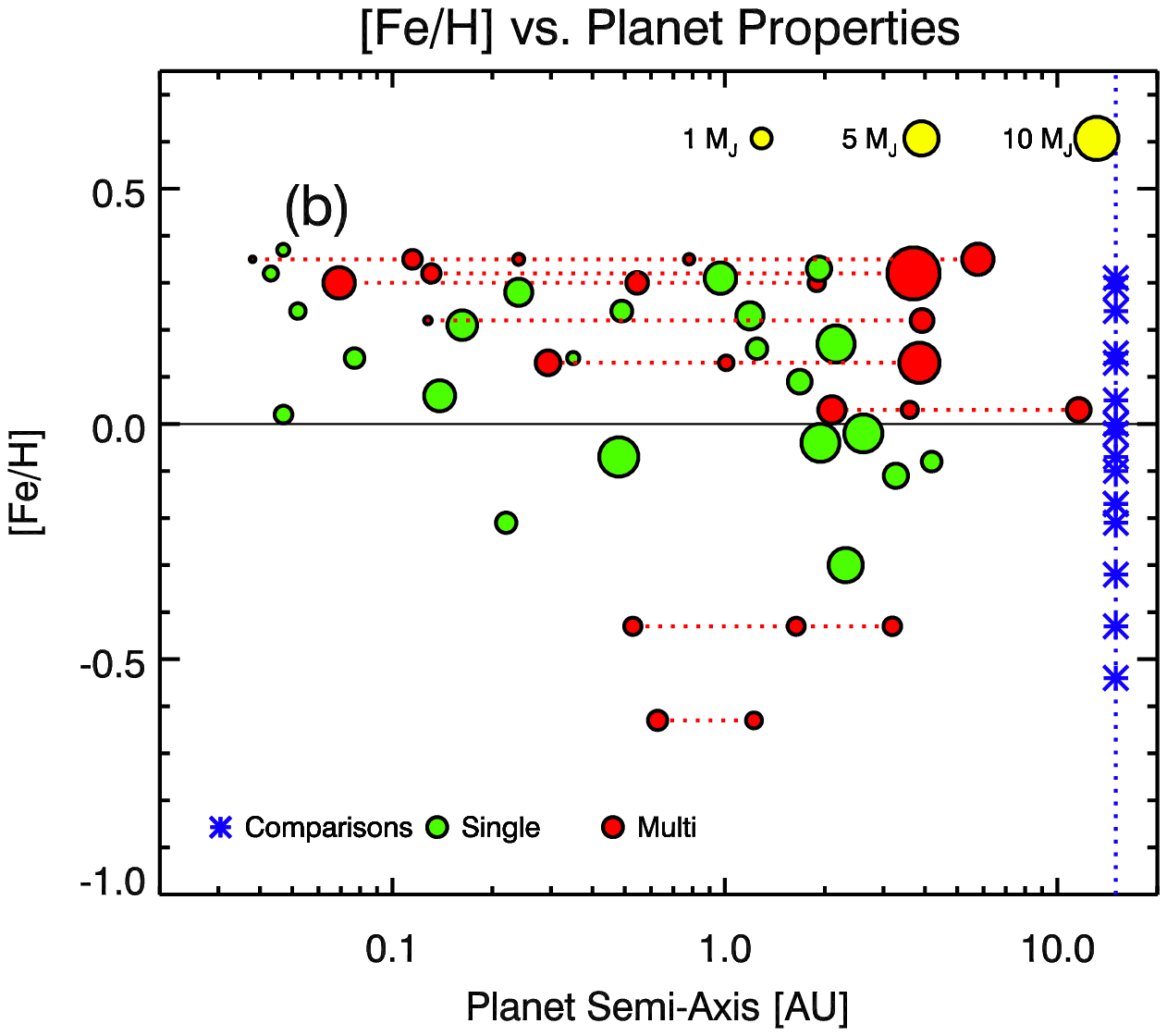} 
\caption{The histogram of metallicity and the relation between metallicity and planet properties. On the $left$ plot, red and blue histogram represent the metallicity distribution of PHSs and comparisons. On the $right$ plot, the $y$ axis is metallicity and the $x$ axis is semi-major axis of planetary orbit. The symbol size stands for the planet mass and the red circles connected with dashed line indicate multiple-planet systems. The blue asterisks in the $right$ plot indicate the metallicity of comparisons.  } 
\label{fig:metal}
\end{figure*}

According to the study for the correlation of planet and metallicity \citep{FV05,JA10}, metallicity is very crucial for the formation of planets, especially, which are massive giants. Although the samples of this study are not fully volume-limited, however, we could confirm that the PHSs tend to be more metal-rich compared to comparisons (\figurename~\ref{fig:metal}a) \citep{SI01,SI03,SI04,FV05,SI05,SS08,JA10}. And the relation between metallicity and planet properties, such as planet mass and semi-major axis of planetary orbit, was examined (\figurename~\ref{fig:metal}b). As shown in \figurename~\ref{fig:metal}b, more massive planets were detected around the PHSs with high [Fe/H] ratio relative to the samples with low [Fe/H] ratio. This relation between planet mass and metallicity have been suggested in previous studies \citep{IL04,FV05}. 
Because it is very unlikely to detect more massive planets in these PHSs by another method of exoplanet detection, it is not expected that this distribution of massive planets will be changed in the future. 

\begin{figure}[!p]
\epsscale{0.8}
\plotone{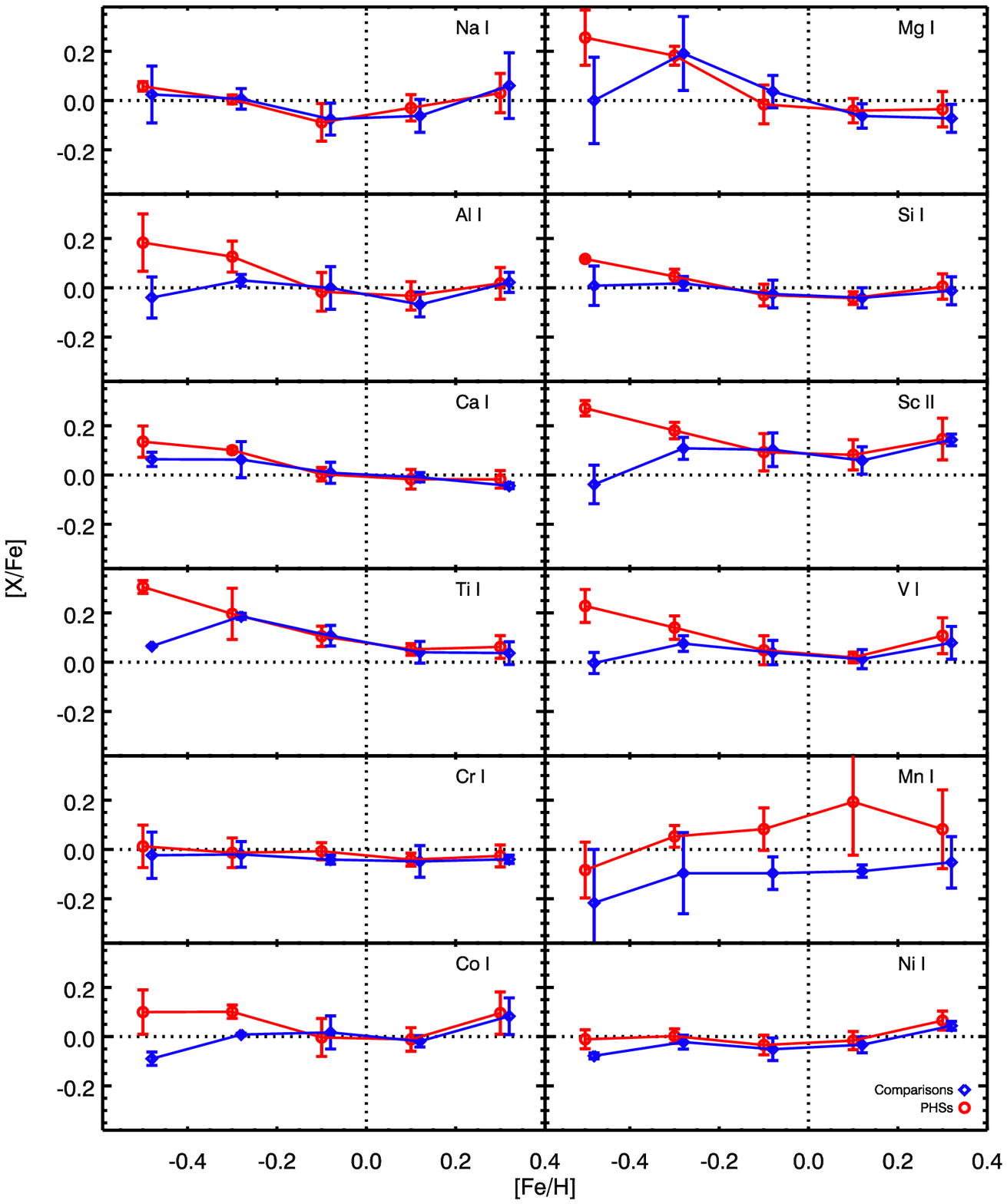} 
\caption{The average and standard deviation of abundances for each [Fe/H] bin, for only dwarfs among the samples. The symbols and error bars respectively indicate the average and standard deviation for each bin. The bins are centered at [Fe/H] = -0.5, -0.3, -0.1, 0.1, 0.3 dex, and the size of bin is 0.2 dex} 
\label{fig:avgsig}
\end{figure}

\subsection{The Average and Standard Deviation of Abundances} 

We carried out the abundance analysis of 12 elements for 52 G-type stars. The difference of [X/Fe] between PHSs and comparison stars was examined to estimate the average of abundances for each element (\figurename~\ref{fig:avgsig}). 
These results for two groups are separately represented for each [Fe/H] bin whose size is 0.2 dex, and the symbols and error bars indicate the average and standard deviation of each group for each [Fe/H] bin. 
Since there is different trend of abundances between metal-rich and metal-poor stars in the Galactic chemical evolution, especially for $\alpha$-elements, it should be necessary to investigate the abundances along the [Fe/H] ratio. 
As shown in \figurename~\ref{fig:avgsig}, most of $\alpha$-elements such as Mg, Ca, Ti of PHSs are more abundant and scattered in [Fe/H] \lt -0.4 relative to comparison stars and [Mn/Fe] shows the most noticeable difference between two groups of PHSs and comparison stars. 
The differences of most elements, such as Mg, Al, Sc, Ti, V, and Co are large as much as about 0.2 dex at [Fe/H] \lt -0.5 and disappear beyond [Fe/H] of -0.3. 

The trend of the [X/Fe] ratio for 12 elements with metallicity is shown in \figurename~\ref{fig:xfe4_all} for two groups of PHSs (red circles) and comparison stars (blue diamonds) separately with background gray symbols of 743 stars by \citet{SG05}. 
And the right plots of each panel indicate abundance of host stars in $y$-axis, semi-major axis of planetary orbit in $x$-axis, and planetary mass as size of symbol.

\subsection{Na, Mg, and Al abundances}

The abundances of Na, Mg and Al elements show similar trend to each other, that decrease with increasing [Fe/H] until the solar metallicity and then slightly increase in the region of [Fe/H] \gt 0 (\figurename~\ref{fig:xfe4_all}). 
The trend for these results has a good agreement with abundance results as gray x-symbols of \citet{SG05} for 743 galactic stars. 
The [Na/Fe] ratios of two groups show remarkable increment in the stars with [Fe/H] \gt 0 and the [Mg/Fe] ratios decrease more steeply than abundances of the other elements in the range [Fe/H] \lt 0. 

But, there are little differences between PHSs and comparisons and no relation between chemical abundances of [Na/Fe], [Mg/Fe], and [Al/Fe] and planet properties. 
These trends of [Na/Fe], [Mg/Fe], and [Al/Fe] in both groups of PHSs and comparison stars have been also shown in the previous results of \citet{GI06} and \citet{NS09}.
The [Mg/Fe] and [Al/Fe] ratios of PHSs are slightly larger than those of comparison stars.
The results of both \citet{GI06} and \citet{NS09} show a similar trend of [Mg/Fe] to our results.
Their [Mg/Fe] ratios of PHSs also seem to be located at the upper boundary of all [Mg/Fe] ratios in low-metallicity samples with [Fe/H] \lt -0.4.  

On the plots with planet properties in \figurename~\ref{fig:xfe4_all}, all Na, Mg, and Al results do not show the noticeable relation between abundances and planet properties. But it is interesting that HD 37124 dwarf, which has three planets (at 0.53, 1.64, 3.19 AU) of about 0.6 $M_J \sin i$, shows very high [Mg/Fe] and [Al/Fe] ratios. 
Since the Al and Mg elements are very important elements in the condensation process \citep{LO03} and abundant in the solar system as much as Fe, though HD 37124 is metal-poor as [Fe/H] = -0.43, the fact that it has as many as these 3 planets could be related with high composition of these Al and Mg elements.

\subsection{Si, Ca, Sc, and Ti Abundances}

The abundances of Si, Ca, Sc, and Ti elements shows typical trends of $\alpha$-elements although [Sc/Fe] shows larger scatter.
The abundances of these elements are increasing with decreasing metallicity. 
The large scatter of [Sc/Fe] among the samples may be caused by hyperfine splitting of \ScII~line.

The abundances of Ti and Sc in the stars with [Fe/H] \lt -0.4 show large difference between PHSs and comparisons. 
Although there are only four stars in [Fe/H] \lt -0.4, the difference of abundance is larger than the typical error of abundance.
This difference in low-metallicity samples has been also found for the other elements, Si and Ca. 
In \figurename~\ref{fig:avgsig}, we can obviously observe that the abundance difference of all these element, such as Si, Ca, Sc, and Ti, become larger in the sample of [Fe/H] \lt -0.4. 
Although \citet{RL06} suggested that [Ti/Fe] was insensitive to the planet occurrence, 
the [X/Fe] ratios of Si, Ca, Sc, and Ti in \citet{NS09} have a good agreement with our results in the range of [Fe/H] \lt -0.4. 
The results of \citet{NS09} also show the high [X/Fe] ratios of Si, Ca, Sc, and Ti for PHSs in low-metallicity samples.
 
The low-metallicity star, HD 155358 ([Fe/H] = -0.63) which has two planets around 0.63 and 1.22 AU has high [Ca/Fe], [Ti/Fe] and [Sc/Fe] ratio. HD 37124 ([Fe/H] = - 0.43) with high [Mg/Fe] and [Al/Fe] ratio also has high [Ti/Fe]. 
It could be caused by intrinsic difference of abundances between thick and thin galactic disk stars. 
We will discuss about the probability that these stars belong to the Galactic thick disk in Sect. 6.
 
\subsection{V, Cr, and Mn abundances}

\begin{figure*}[!p]
\epsscale{1}
\plottwo{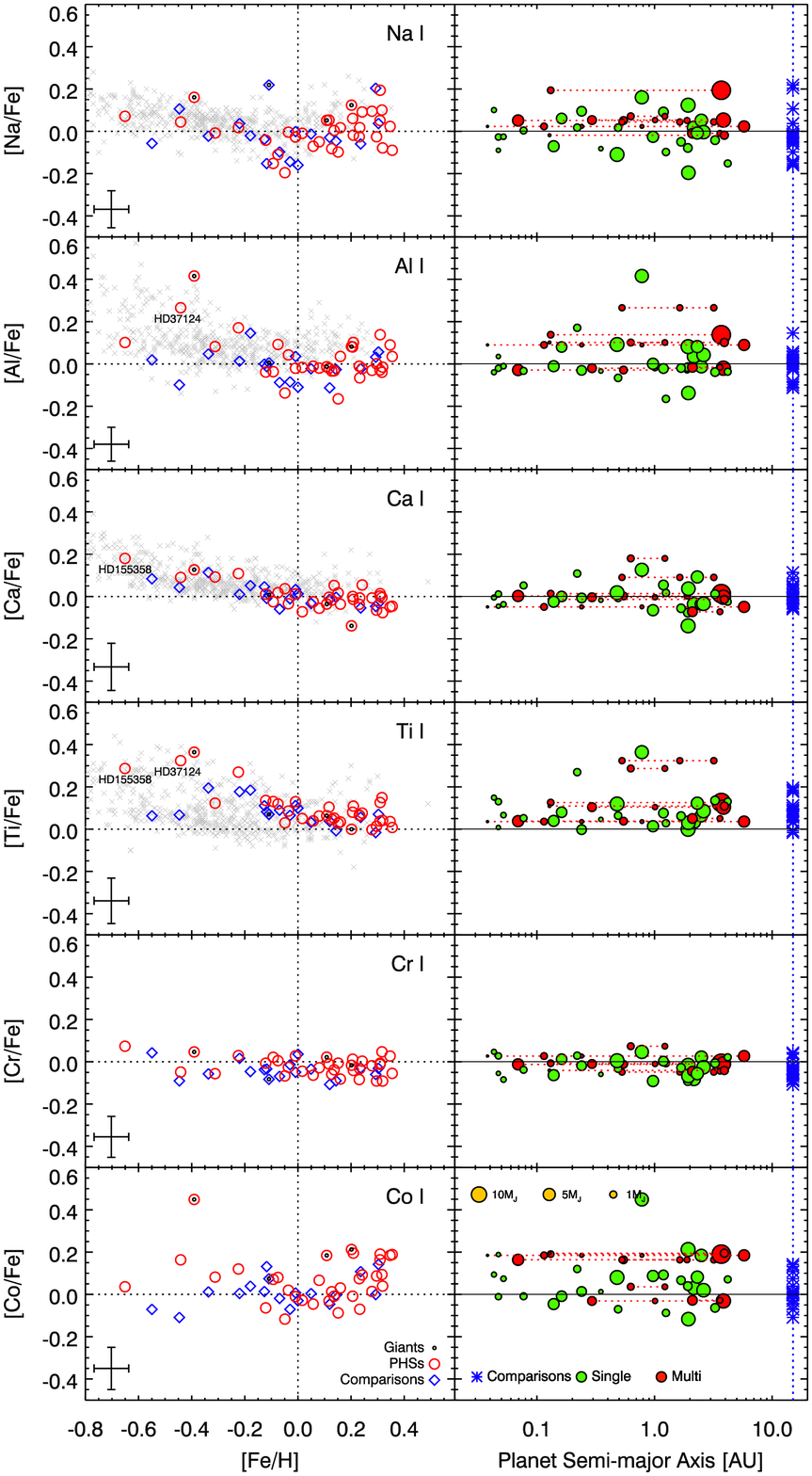}{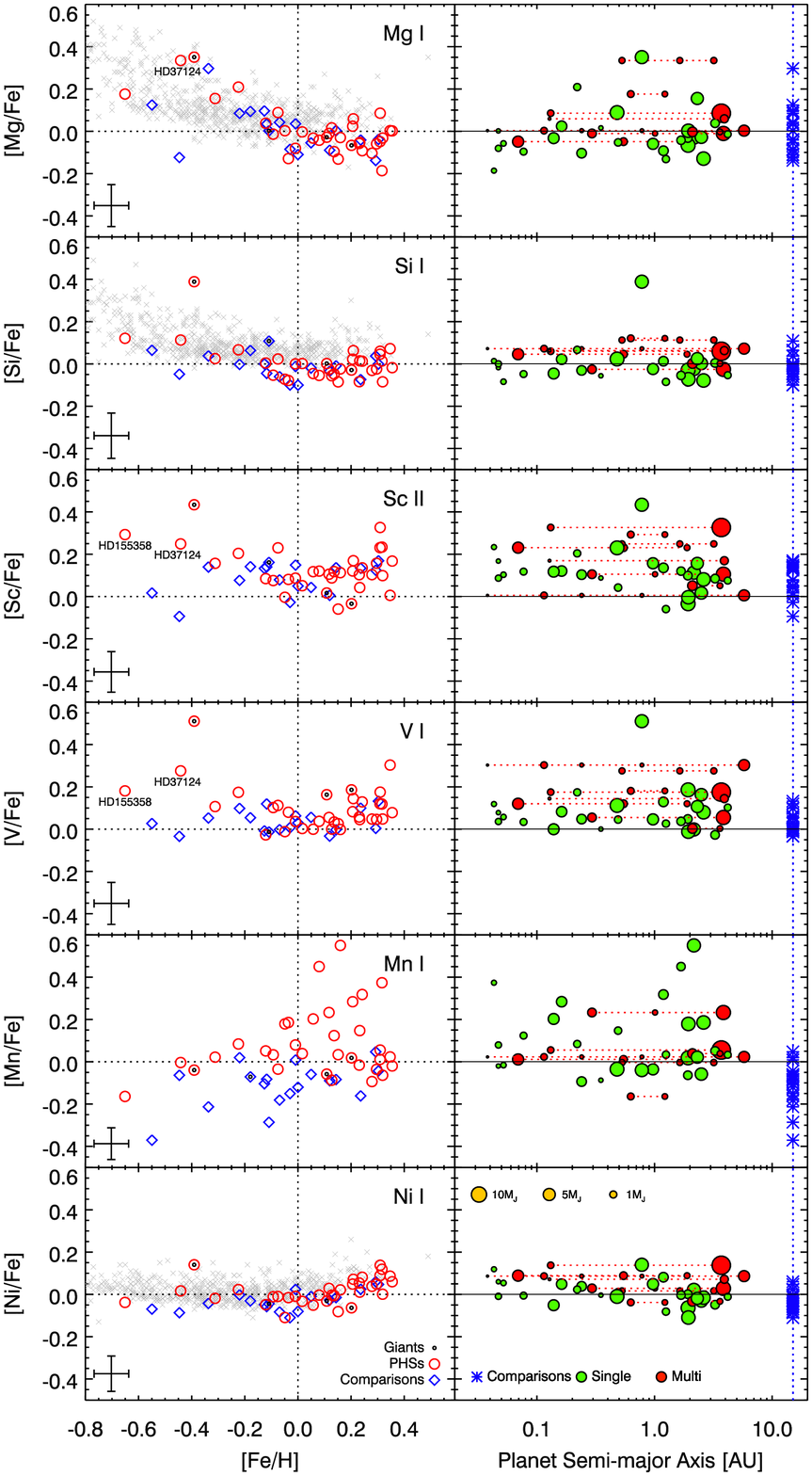} 
\caption{The elemental abundance results: [X/Fe] vs. [Fe/H] ($left$) and [X/Fe] vs. planet's properties ($right$). 
The $left$ plots of each panel show the trends of 12 elements abundances [X/Fe] with [Fe/H]. 
The cross hair at the left-bottom of each panel means typical errors of abundances as the average of errors, and the red circles and blue diamonds represent PHSs and comparisons, and the symbols including black dot mean the giant stars. 
The gray x-symbols indicate the results of \citet{SG05} for 743 stars. 
The $right$ of each panel demonstrates the relation between elemental abundances and planet properties. 
The symbols of planets are the same to those in the plots of \figurename~\ref{fig:targets}.
} 
\label{fig:xfe4_all}
\end{figure*}

The trend of V, Cr, and Mn abundances in \figurename~\ref{fig:xfe4_all} seems different from those of $\alpha$-elements.
The abundances of V show large scatter compared to Cr abundances among all target stars. 
The Cr and V are $iron$-peak elements which have the same origin with Fe in the nucleosynthesis process. 
Despite of large scatter for each star, however, the V abundances also show slight difference between PHSs and comparisons in the stars of [Fe/H] \lt 0, like $\alpha$-elements as shown in [Ti/Fe] and [Sc/Fe] vs. [Fe/H] plots of \figurename~\ref{fig:xfe4_all}, but not for Cr abundance.
The low-metallicity star, HD 37124 shows high [V/Fe] ratio similar to high [Mg/Fe] and [Al/Fe] ratios. 

For Mn abundance, the trend of [Mn/Fe] with respect to [Fe/H] is different from those of $\alpha$-elements and $iron$-peak elements. 
The difference between PHSs and comparisons is over 0.15 dex on whole range of [Fe/H]. 
On the other hand, the Mn abundances estimated by G06 show also very scattered like our results, but the result of G06 do not show noticeable difference of [Mn/Fe] between PHSs and comparisons. 
It is very difficult to measure the accurate EWs of \MnI~lines due to the contamination of nearby lines and hyperfine splitting, so the Mn abundances were determined by a few lines, which were obviously isolated and well fitted by Gaussian profile. 
Even taken into account these scatters from measurements of a few lines, the difference of [Mn/Fe] between PHSs and comparisons seems not to be due to only errors in determination of abundance. 
Furthermore, other previous studies have presented this discrepancy of [Mn/Fe] between PHSs and comparisons \citep{BS03,ZC02}.
This may imply that Mn element could be somehow related with the process of planet formation. 

\subsection{Co and Ni abundances}

 The abundance trends of Co and Ni are similar to Fe abundances, [Fe/H]. But the [Co/Fe] ratios show large scatter compared with [Ni/Fe] ratios. In the [Co/Fe] vs. [Fe/H] plot, the [Co/Fe] ratios of PHSs are also slightly higher than those of comparison samples in the sample of [Fe/H] \lt 0. 

Although the difference of [Co/Fe] and [Ni/Fe] in the range of [Fe/H] \lt 0 is small relative to those of the other elements, our trends of [Co/Fe] and [Ni/Fe] in low-metallicity stars show a good agreement with those of \citet{NS09}.  
The abundances of Co and Ni for stars with planets in \citet{NS09} also show slightly higher [X/Fe] ratios than the average of [X/Fe] ratios for stars without planets around [Fe/H] = -0.4.

\subsection{Comparison of the Abundances with the Reference}

\begin{figure}[!t]
\epsscale{1}
\plotone{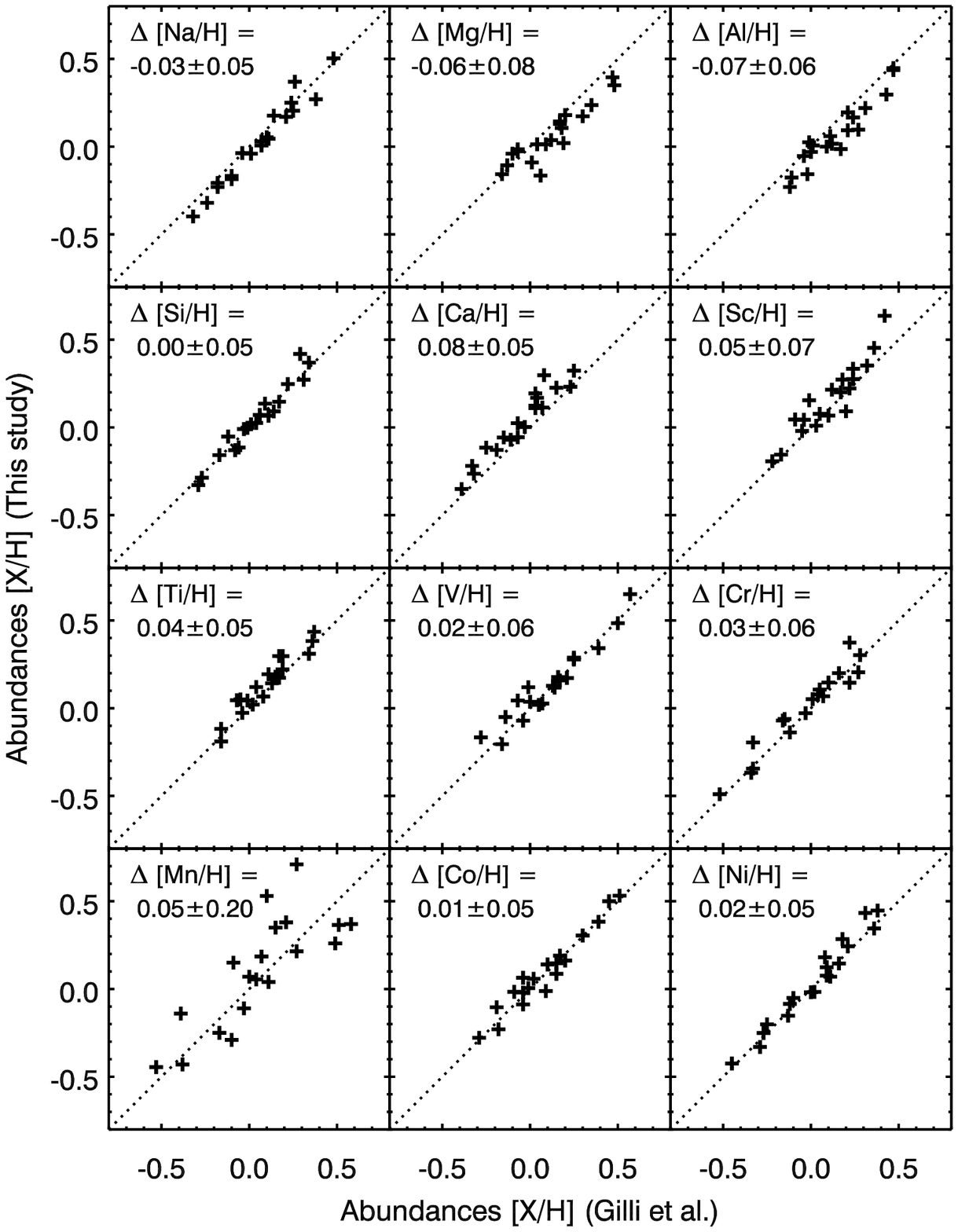}
\caption{The abundance difference between this work and Gilli et al.(2006) The average and standard deviation in the same samples are indicated in the right side of plot.} 
\label{fig:compabd}
\end{figure}

To test how reliable these results are, we compared these abundances with the results adopted from Gilli et al. (2006, hereafter G06). 
There are the 19 same samples with G06 and \figurename~\ref{fig:compabd} shows that our results of elemental abundance, especially [Fe/H] and [Si/H], have a good agreement with the results of G06. 
Except Mn, our abundances of most elements are consistent to the abundances of G06 within 0.08 dex of standard deviation. 
The Mn abundance in this work shows a large scatter with those of G06 relative to the other elements. 
Since \MnI\ lines have a hyperfine structure, these scatters should be examined in detail using hyperfine structure analysis. 

\section{Discussion}

\subsection{Abundance Differences between PHSs and Comparisons}

\begin{deluxetable}{rrcrrcrcrr} 
\tablecaption{The average and standard deviation of abundances} 
\tabletypesize{\footnotesize}
\tablewidth{0cm}
\tablehead{
& \multicolumn{4}{c}{for 16 dwarfs with [Fe/H] \lt 0 } &  &
  \multicolumn{4}{c}{for 32 dwarfs with [Fe/H] \gt 0 } \\[5pt]
\cline{2-5} \cline{7-10} \\[-5pt]
\colhead{Elements} &  
\colhead{$\mathrm{[X/Fe]_{PHS}}$} & 
\colhead{} &  
\colhead{$\mathrm{[X/Fe]_{Comp}}$} & 
\colhead{$\mathrm{\Delta[X/Fe]}$}\tablenotemark{a} &
\colhead{} &
\colhead{$\mathrm{[X/Fe]_{PHS}}$} & 
\colhead{} &  
\colhead{$\mathrm{[X/Fe]_{Comp}}$} & 
\colhead{$\mathrm{\Delta[X/Fe]}$}\tablenotemark{a} 
}
\startdata 
    Na I & -0.04 $\pm$ 0.09 & & -0.04 $\pm$ 0.08 &    0.00  & &  0.01 $\pm$ 0.08 & & -0.01 $\pm$ 0.11 &    0.02 \\
    Mg I &  0.08 $\pm$ 0.14 & &  0.06 $\pm$ 0.12 &    0.02  & & -0.04 $\pm$ 0.06 & & -0.07 $\pm$ 0.05 &    0.03 \\
    Al I &  0.05 $\pm$ 0.12 & & -0.00 $\pm$ 0.07 &    0.05  & & -0.00 $\pm$ 0.07 & & -0.03 $\pm$ 0.06 &    0.03 \\
    Si I &  0.01 $\pm$ 0.07 & & -0.01 $\pm$ 0.05 &    0.03  & & -0.01 $\pm$ 0.05 & & -0.03 $\pm$ 0.05 &    0.02 \\
    Ca I &  0.05 $\pm$ 0.07 & &  0.03 $\pm$ 0.05 &    0.02  & & -0.02 $\pm$ 0.04 & & -0.02 $\pm$ 0.02 &    0.01 \\
   Sc II &  0.15 $\pm$ 0.10 & &  0.08 $\pm$ 0.08 &    0.07  & &  0.12 $\pm$ 0.08 & &  0.09 $\pm$ 0.06 &    0.03 \\
    Ti I &  0.16 $\pm$ 0.10 & &  0.11 $\pm$ 0.05 &    0.05  & &  0.06 $\pm$ 0.04 & &  0.04 $\pm$ 0.04 &    0.02 \\
     V I &  0.10 $\pm$ 0.09 & &  0.04 $\pm$ 0.05 &    0.07  & &  0.07 $\pm$ 0.07 & &  0.04 $\pm$ 0.06 &    0.03 \\
    Cr I & -0.00 $\pm$ 0.04 & & -0.03 $\pm$ 0.04 &    0.03  & & -0.03 $\pm$ 0.04 & & -0.05 $\pm$ 0.05 &    0.01 \\
    Mn I &  0.04 $\pm$ 0.10 & & -0.12 $\pm$ 0.12 &    0.16  & &  0.13 $\pm$ 0.19 & & -0.07 $\pm$ 0.07 &    0.20 \\
    Co I &  0.04 $\pm$ 0.08 & & -0.01 $\pm$ 0.07 &    0.04  & &  0.05 $\pm$ 0.09 & &  0.02 $\pm$ 0.07 &    0.03 \\
    Ni I & -0.02 $\pm$ 0.04 & & -0.05 $\pm$ 0.04 &    0.03  & &  0.03 $\pm$ 0.05 & & -0.00 $\pm$ 0.05 &    0.03 \\
\enddata
\tablenotetext{a}{$\mathrm{<[Fe/H]>_{PHSs}-<[Fe/H]>_{Comparisons}}$}
\label{tab:avgsig} 
\end{deluxetable}

We have taken the two groups of samples, PHSs and comparison stars (non-planet host stars). 
So the average of abundance was compared between two groups for only dwarf samples, in order to eliminate the effects due to convection near the surface. 
In \tablename~\ref{tab:avgsig}, the averages and standard deviations of abundances are shown for dwarf stars in different metallicity range, the stars with [Fe/H] \lt 0 and those with [Fe/H] \gt 0. 
In the case of most elements, the abundances of PHSs are slightly larger than those of comparisons, although the differences are tiny, relative to standard deviation of samples. 
For most elements, the abundance differences between PHSs and comparison stars of [Fe/H] \lt 0 are greater than those of [Fe/H] \gt 0. 
Thus, we confirmed that differences of elemental abundances are more outstanding in stars with [Fe/H] \lt 0. It would become more significant in searching for Earth-class exoplanets, because previous studies of elemental abundances related with planet-host stars have suggested that it is more feasible to find these low-mass planets in the low-metallicity stars \citep{UM06,SS08}. 
In \figurename~\ref{fig:xfe4_all}, we have already examined that the sample, HD 37124 with low metallicity and high [Mg/Fe], [Al/Fe], [Ti/Fe], and [V/Fe] has three planets of relatively low-mass ($<~1~M_\mathrm{J}$) and HD 155358 with high [Ca/Fe], [Sc/Fe], and [Ti/Fe] has two relatively low-mass planets.    .

\subsection{Kolmogorov-Smirnov Test}   

\begin{figure*}[!htp]
\centering{
	\includegraphics[width=0.30\textwidth]{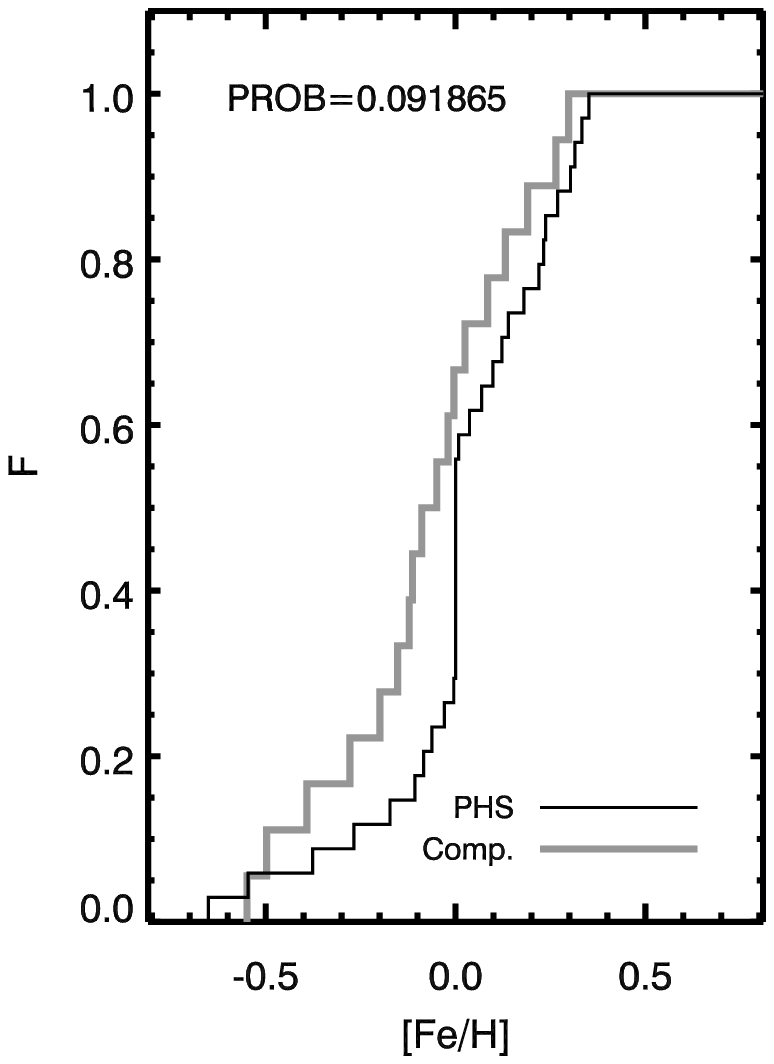}
	\includegraphics[width=0.67\textwidth]{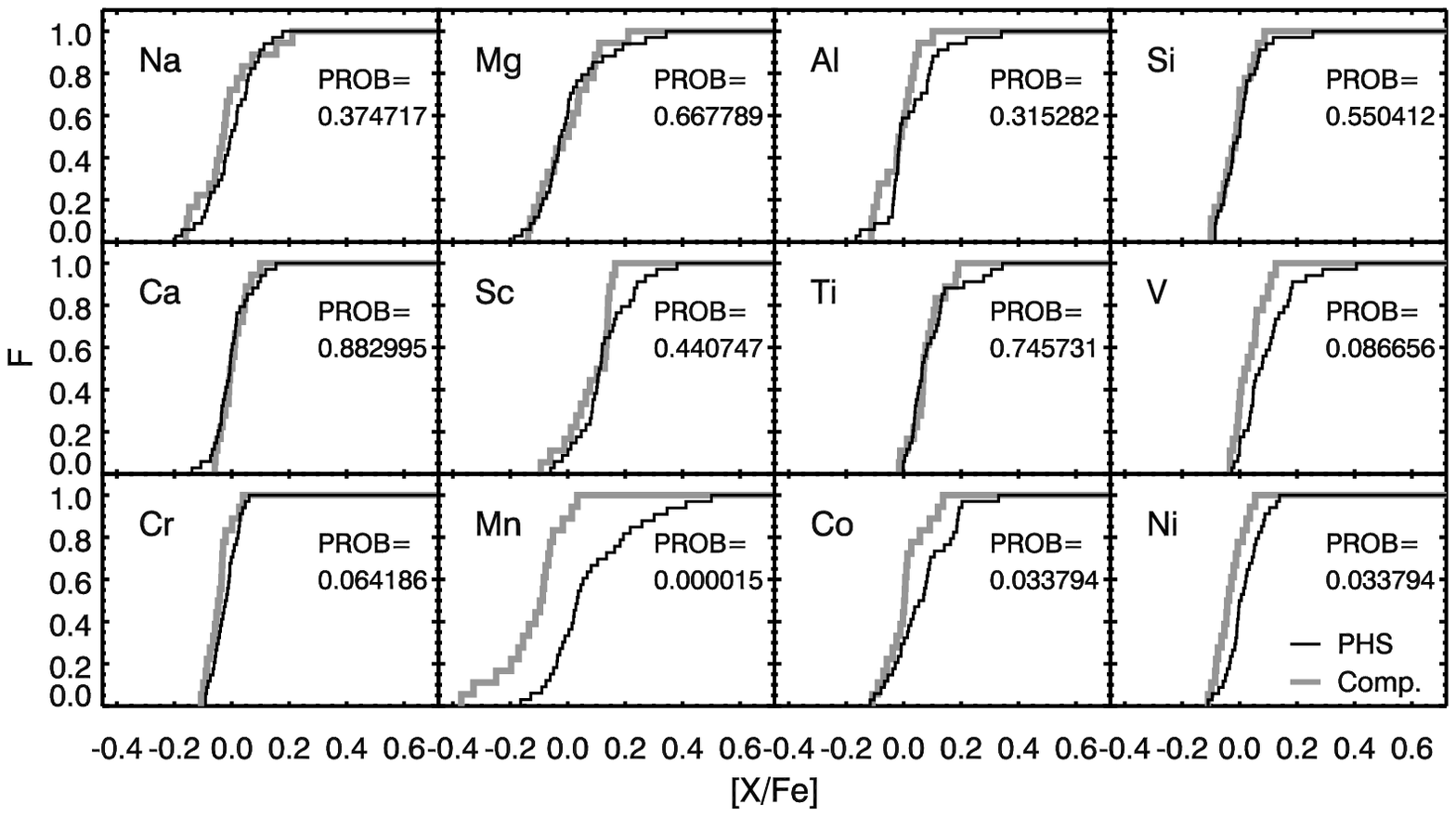}
}
\caption{The cumulative distribution function for [Fe/H] and other elements. The $x$-axis is [Fe/H] and [X/Fe] of elemental abundances and the $y$-axis is cumulative fraction of samples. The value of $PROB$ shows the significance level of the K-S statistics, and in the case of smaller $PROB$ value, the two data sets would have more different distribution each other.} 
\label{fig:cdf}
\end{figure*}

In order to evaluate the probability that abundances of two groups have the same distribution, Kolmogorov-Smirnov test (K-S test) was performed between two groups of PHSs and comparison stars. 
Since the K-S test is available to investigate the difference of [X/Fe] distribution in the case of small size of sample, it is useful to compare the abundance distribution of PHS with that of comparison.
The cumulative distribution function for [Fe/H] and other elements are shown in \figurename~\ref{fig:cdf}. 
In \figurename~\ref{fig:cdf}, the $x$-axis is [Fe/H] or [X/Fe] of elemental abundances and the $y$-axis is cumulative fraction of samples. 
The value of $PROB$ represents the significance level of the K-S statistics as the probability that two distributions belong to the same population.
The small $PROB$ value means that the two groups of data have different distribution from each other.

For metallicity distribution, the probability that the [Fe/H] distribution of two groups belongs to the same population was about 9\%. 
As shown in the cumulative distribution function of [Mn/Fe] of \figurename~\ref{fig:cdf}, the probability for [Mn/Fe] distribution was about 0.0015\%, and it was only thousandth of those for other elements. 
Although the size of sample was small and there are random errors of abundances, this result shows that the distribution of [Mn/Fe] for each group was severely different from each other. 
For Cr and Ni elements, the [X/Fe] of PHS and comparison shows the similar trend in the whole range of [Fe/H] in the left plots of \figurename~\ref{fig:xfe4_all}. 
Despite of that, the probability that the [X/Fe] distributions of Cr and Ni in two groups belong to the same population was only about 7\%, 3\%, respectively and smaller than that for metallicity.
For the other elements, the probabilities of [X/Fe] distribution are larger than that of [Fe/H] distribution.
 
\section{Concluding Remark}

We have carried out abundance analysis for 12 elements in the samples of 34 PHSs and 18 comparison stars of only G-type stars. 
Since it would be expected that the differences of abundances among samples was small as about 0.1 dex, we concentrated on minimizing the systematic errors. 
We limited the samples to G-type stars that were similar to the Sun, and the entire processes of abundance analysis were restrictively performed in the uniform way. 
Using the abundances and planet properties of PHSs, such as planet mass and semi-major axis, we have been plotted the ratios [X/Fe] vs. [Fe/H] and [X/Fe] vs. semi-major axis of planetary orbit with planet mass. 
In the plot of [Fe/H] and planet properties, we could confirm that host stars with low metallicity tended to bear less massive multiple planets. 

\begin{figure}[!htp]
\plotone{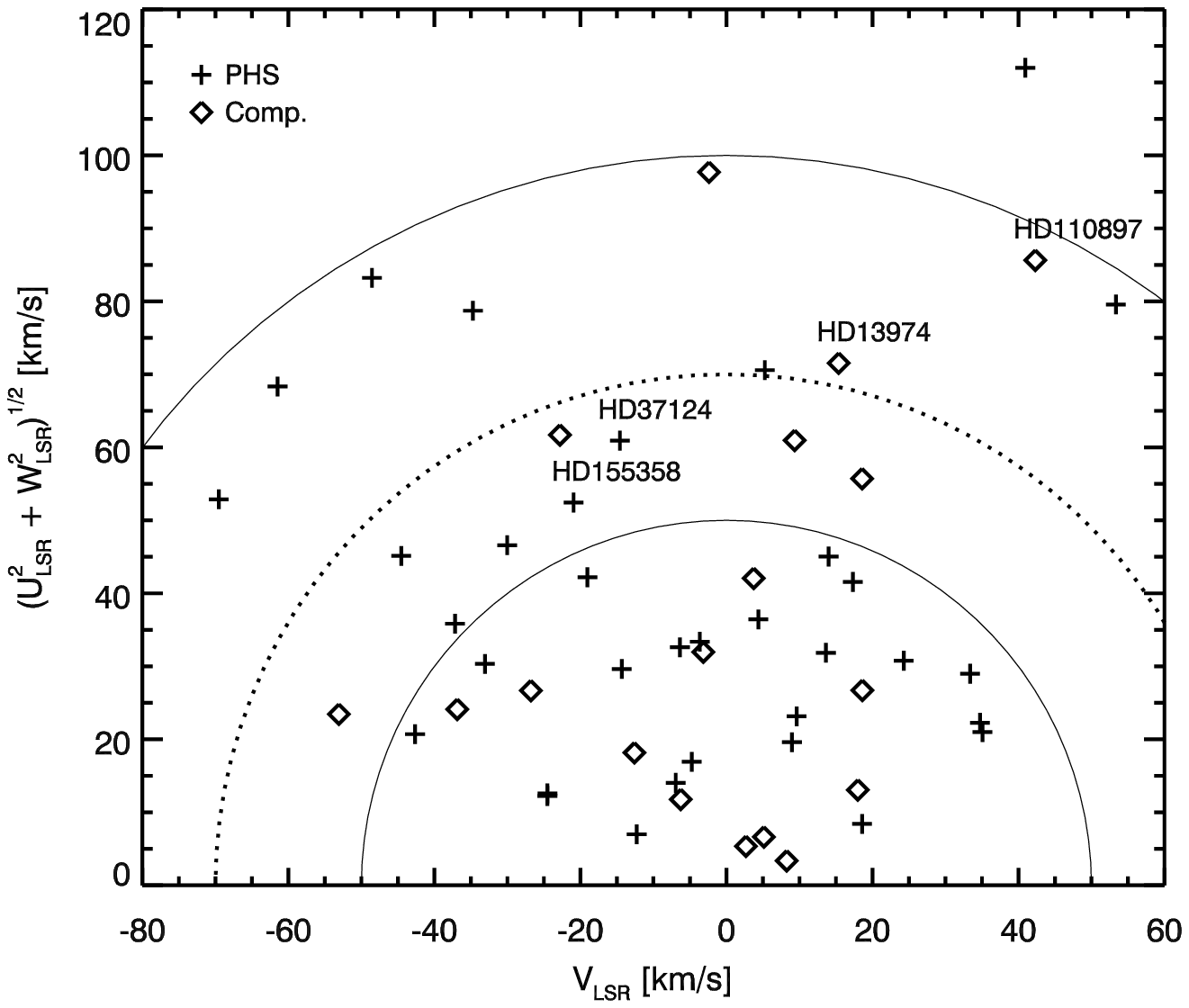}
\caption{The Toomre diagram for U, V, and W velocities of our samples. The U, V, and W velocities were calculated from our radial velocities and the proper motions, which are adopted from SIMBAD. This plot shows the Galactic velocities of PHSs (HD 37124 and HD 155358) and comparison stars (HD 13974 and HD 110897) with [Fe/H] \lt -0.4. Their V components of four samples are larger than $-50$ \kms. In the case of low-metallicity PHSs, their peculiar space velocities, $v_{pec} = ( U_{LSR}^2 + V_{LSR}^2 + W_{LSR}^2)^{1/2}$ are also smaller than 70 \kms. \label{fig:uvw}}
\end{figure}

In previous studies, the authors have made every effort to investigate the statistical difference of abundances only between PHSs and normal field stars in whole range of metallicity. 
The chemical anomalies of PHSs, however, would be noticeable in the stars with low metallicity, because the total amount of metal was insufficient so that those planets were in a hard situation to be formed.
In \figurename~\ref{fig:avgsig}, the [X/Fe] of most elements have shown slight discrepancy between two groups in the stars of [Fe/H] \lt -0.4. 
The [X/Fe] of Mg, Al, Sc, Ti, V, and Co for PHSs are higher than those of comparison stars in the region of [Fe/H] \lt -0.4 by more than 0.2 dex. 
HD 37124 (3 planets at 0.53, 1.64, 3.19 AU) and HD 155358 (2 planets at 0.63, 1.22 AU), even though they are poor in metal, have several planets with the mass less than 1 $M_\mathrm{J}$. 
Although there are two samples of PHSs in the range of [Fe/H] \lt -0.4 as shown in \figurename~\ref{fig:xfe4_all}, these PHSs imply that the PHSs with low metallicity and with high [Mg/Fe], [Al/Fe], [Ca/Fe], [Sc/Fe], [Ti/Fe], and [V/Fe] tend to bear the low-mass planets, while the metal-rich PHSs tend to have massive planets. 
When these stars were verified by U, V, and W velocities and location in our Galaxy, both two stars are located near the Sun and the Galactic plane, and their velocities are small relative to those of thick disk stars in \figurename~\ref{fig:uvw} \citep{BF03}. 
So it is unlikely for these PHSs to belong to the Galactic thick disk, rather than for two comparison stars.    

In this study, we have found that Mn is the most interesting element. 
As mentioned above, the [Mn/Fe] ratios of most PHSs are higher than those of comparisons as much as about 0.15 dex and the result of K-S test implies that the distribution of [Mn/Fe] shows that the [Mn/Fe] ratios in two groups have the different population . 
Although the Mn abundances were obtained from a few lines, this large discrepancy of [Mn/Fe] ratio is unlikely to come from the errors of abundance analysis. 
And previous studies \citep{BS03,ZC02} have suggested the difference of [Mn/Fe] between PHSs and comparisons, though the statistically within their error range. 

Furthermore, the condensation temperature of Mn is relatively lower than other elements. Then, in the processes of planet formation, the Mn elements would be condensed later than other elements, where the dust ball had been already formed and planetesimals began to form. Therefore, after the compounds of Ca, Ti, Mg, Al, Si elements were firstly formed, the Mn element was condensed into their compounds so that, even though the amount was tiny, we might guess that Mn element played a critical role for condensation process in this stage like a enzyme in chemical reactions. 

In the future, we plan to perform abundance analysis for other elements than refractory and extend the samples to F and K type stars. Thus, we expect to follow up a clue for the presence of exoplanets using other elements, and more sample stars.



\acknowledgements
This work was supported by Human Resources Development Program through the National Research Foundation of Korea funded by the Ministry of Education, Science and Technology.

\clearpage

\input{table4}
\input{table5}

\input{table6}

\end{document}

%% file: table4.tex
\begin{deluxetable}{llrrrrr} 
\tabletypesize{\scriptsize}
\tablecaption{Line list}
\tablewidth{0cm}
\tablehead{
\colhead{$\lambda$} &
\colhead{Element} & 
\colhead{E.P.} & 
\colhead{log $gf$} &
\colhead{EW$_\odot$} & 
\colhead{log $\epsilon_\odot$} & 
}
\startdata

\multicolumn{7}{l}{\textbf{Na I  A$_\odot$ =  6.27}}   \\
 5688.19 & Na I &  2.10 & -0.42 &  134.4 &    6.22 &      \\
 6154.23 & Na I &  2.10 & -1.53 &   38.3 &    6.26 &      \\
 6160.75 & Na I &  2.10 & -1.32 &   57.0 &    6.32 &      \\
\multicolumn{7}{l}{\textbf{Mg I  A$_\odot$ =  7.60}}   \\
 5711.09 & Mg I &  4.34 & -1.72 &  116.4 &    7.58 &      \\
 6319.24 & Mg I &  5.11 & -2.32 &   26.7 &    7.62 &      \\
\multicolumn{7}{l}{\textbf{Al I  A$_\odot$ =  6.50}}   \\
 6696.03 & Al I &  3.14 & -1.57 &   38.1 &    6.50 &      \\
 6698.67 & Al I &  3.14 & -1.88 &   21.6 &    6.48 &      \\
 7836.13 & Al I &  4.02 & -0.56 &   61.7 &    6.51 &      \\
\multicolumn{7}{l}{\textbf{Si I  A$_\odot$ =  7.61}}   \\
 5684.48 & Si I &  4.95 & -1.73 &   67.8 &    7.70 &      \\
 5701.10 & Si I &  4.93 & -2.05 &   39.8 &    7.57 &      \\
 5772.15 & Si I &  5.08 & -1.75 &   57.5 &    7.68 &      \\
 5948.54 & Si I &  5.08 & -1.23 &   96.1 &    7.67 &      \\
 6125.02 & Si I &  5.61 & -1.46 &   36.9 &    7.52 &      \\
 6131.85 & Si I &  5.62 & -1.62 &   26.3 &    7.47 &      \\
 6145.02 & Si I &  5.61 & -1.44 &   43.6 &    7.59 &      \\
 6243.81 & Si I &  5.62 & -1.24 &   50.0 &    7.50 &      \\
 7003.57 & Si I &  5.96 & -0.83 &   69.3 &    7.57 &      \\
 7034.90 & Si I &  5.87 & -0.88 &   74.3 &    7.63 &      \\
 7405.77 & Si I &  5.61 & -0.82 &  104.4 &    7.71 &      \\
 7415.95 & Si I &  5.61 & -0.73 &  105.3 &    7.63 &      \\
 7423.50 & Si I &  5.62 & -0.58 &  116.5 &    7.59 &      \\
\multicolumn{7}{l}{\textbf{Ca I  A$_\odot$ =  6.23}}   \\
 5512.99 & Ca I &  2.93 & -0.30 &   95.2 &    6.23 &      \\
 5581.97 & Ca I &  2.52 & -0.56 &  103.2 &    6.30 &      \\
 5590.11 & Ca I &  2.52 & -0.71 &   92.7 &    6.31 &      \\
 5857.45 & Ca I &  2.93 &  0.23 &  153.9 &    6.28 &      \\
 5867.56 & Ca I &  2.93 & -1.57 &   24.1 &    6.32 &      \\
 6161.30 & Ca I &  2.52 & -1.03 &   67.9 &    6.21 &      \\
 6166.44 & Ca I &  2.52 & -0.90 &   73.7 &    6.18 &      \\
 6169.04 & Ca I &  2.52 & -0.54 &  100.7 &    6.22 &      \\
 6169.56 & Ca I &  2.52 & -0.27 &  123.1 &    6.23 &      \\
 6449.82 & Ca I &  2.52 & -0.50 &   99.9 &    6.16 &      \\
 6455.60 & Ca I &  2.52 & -1.34 &   56.3 &    6.31 &      \\
 6471.66 & Ca I &  2.52 & -0.59 &   94.7 &    6.17 &      \\
 6499.65 & Ca I &  2.54 & -0.59 &   91.0 &    6.14 &      \\
 7148.15 & Ca I &  2.71 &  0.22 &  159.3 &    6.16 &      \\
\multicolumn{7}{l}{\textbf{Sc II  A$_\odot$ =  3.16}}  \\
 5526.79 &Sc II &  1.77 &  0.13 &   80.0 &    3.13 &      \\
 5684.20 &Sc II &  1.51 & -1.08 &   36.0 &    3.15 &      \\
 6245.64 &Sc II &  1.51 & -1.13 &   33.9 &    3.13 &      \\
 6604.60 &Sc II &  1.36 & -1.31 &   37.7 &    3.23 &      \\
\multicolumn{7}{l}{\textbf{Ti I  A$_\odot$ =  4.94}}   \\
 5016.16 & Ti I &  0.85 & -0.57 &   63.1 &    4.94 &      \\
 5022.87 & Ti I &  0.83 & -0.43 &   72.1 &    4.98 &      \\
 5036.46 & Ti I &  1.44 &  0.13 &   70.1 &    4.95 &      \\
 5039.96 & Ti I &  0.02 & -1.13 &   71.4 &    4.87 &      \\
 5043.58 & Ti I &  0.84 & -1.73 &   12.9 &    4.89 &      \\
 5194.04 & Ti I &  2.10 & -0.65 &   11.0 &    4.94 &      \\
 5210.39 & Ti I &  0.05 & -0.88 &   83.7 &    4.92 &      \\
 5219.70 & Ti I &  0.02 & -2.29 &   26.3 &    5.03 &      \\
 5282.38 & Ti I &  1.05 & -1.30 &   24.8 &    5.01 &      \\
 5426.26 & Ti I &  0.02 & -3.01 &    6.2 &    4.98 &      \\
 5471.20 & Ti I &  1.44 & -1.39 &    8.1 &    4.88 &      \\
 5490.15 & Ti I &  1.46 & -0.93 &   19.0 &    4.87 &      \\
 5648.57 & Ti I &  2.49 & -0.25 &   10.1 &    4.85 &      \\
 5662.16 & Ti I &  2.32 & -0.11 &   22.2 &    4.96 &      \\
 5689.49 & Ti I &  2.30 & -0.47 &   13.6 &    5.03 &      \\
 5739.98 & Ti I &  2.24 & -0.67 &    7.3 &    4.87 &      \\
 5880.27 & Ti I &  1.05 & -2.05 &    4.7 &    4.88 &      \\
 5903.32 & Ti I &  1.07 & -2.14 &    3.7 &    4.88 &      \\
 5922.11 & Ti I &  1.05 & -1.47 &   18.1 &    4.95 &      \\
 5978.54 & Ti I &  1.87 & -0.50 &   21.5 &    4.88 &      \\
 6064.63 & Ti I &  1.05 & -1.94 &    7.9 &    5.00 &      \\
 6091.17 & Ti I &  2.27 & -0.42 &   14.7 &    4.98 &      \\
 6126.22 & Ti I &  1.07 & -1.42 &   22.1 &    5.03 &      \\
 6258.10 & Ti I &  1.44 & -0.35 &   51.1 &    4.94 &      \\
 6261.10 & Ti I &  1.43 & -0.48 &   49.1 &    5.01 &      \\
 6303.76 & Ti I &  1.44 & -1.57 &    7.2 &    4.95 &      \\
 6312.24 & Ti I &  1.46 & -1.55 &    6.9 &    4.94 &      \\
 6556.06 & Ti I &  1.46 & -1.07 &   16.8 &    4.89 &      \\
 7138.90 & Ti I &  1.44 & -1.59 &    6.5 &    4.88 &      \\
\multicolumn{7}{l}{\textbf{Ti II  A$_\odot$ =  4.92}}  \\
 4708.66 &Ti II &  1.24 & -2.34 &   52.0 &    4.95 &      \\
 5005.16 &Ti II &  1.57 & -2.72 &   21.2 &    4.91 &      \\
 5185.91 &Ti II &  1.89 & -1.46 &   65.2 &    4.95 &      \\
 5336.77 &Ti II &  1.58 & -1.59 &   74.5 &    4.98 &      \\
\multicolumn{7}{l}{\textbf{V  I  A$_\odot$ =  3.90}}   \\
 5668.37 & V  I &  1.08 & -1.03 &    5.3 &    3.95 &      \\
 5670.85 & V  I &  1.08 & -0.43 &   16.9 &    3.92 &      \\
 5727.05 & V  I &  1.08 & -0.01 &   36.5 &    3.98 &      \\
 5727.66 & V  I &  1.05 & -0.87 &    7.2 &    3.90 &      \\
 5737.07 & V  I &  1.06 & -0.74 &    9.0 &    3.89 &      \\
 6081.44 & V  I &  1.05 & -0.58 &   13.7 &    3.91 &      \\
 6090.22 & V  I &  1.08 & -0.06 &   33.5 &    3.95 &      \\
 6119.52 & V  I &  1.06 & -0.32 &   21.2 &    3.90 &      \\
 6199.20 & V  I &  0.29 & -1.30 &   12.3 &    3.81 &      \\
 6243.10 & V  I &  0.30 & -0.98 &   26.0 &    3.92 &      \\
 6251.82 & V  I &  0.29 & -1.34 &   12.8 &    3.87 &      \\
 6274.64 & V  I &  0.27 & -1.67 &    7.1 &    3.89 &      \\
 6285.16 & V  I &  0.28 & -1.51 &    9.0 &    3.85 &      \\
\multicolumn{7}{l}{\textbf{Cr I  A$_\odot$ =  5.65}}   \\
 5304.18 & Cr I &  3.46 & -0.69 &   15.5 &    5.63 &      \\
 5312.86 & Cr I &  3.45 & -0.56 &   18.2 &    5.57 &      \\
 5318.77 & Cr I &  3.44 & -0.69 &   15.0 &    5.58 &      \\
 5783.07 & Cr I &  3.32 & -0.50 &   31.6 &    5.71 &      \\
 5787.92 & Cr I &  3.32 & -0.08 &   46.9 &    5.59 &      \\
 6979.80 & Cr I &  3.46 & -0.41 &   37.1 &    5.81 &      \\
\multicolumn{7}{l}{\textbf{Mn I  A$_\odot$ =  5.33}}   \\
 4265.92 & Mn I &  2.94 & -0.27 &   63.0 &    5.31 &      \\
 6440.93 & Mn I &  3.77 & -1.24 &    5.7 &    5.35 &      \\
\multicolumn{7}{l}{\textbf{Fe I  A$_\odot$ =  7.53}}   \\
 4439.89 & Fe I &  2.28 & -3.00 &   49.6 &    7.44 &      \\
 4523.40 & Fe I &  3.65 & -1.99 &   41.6 &    7.52 &      \\
 4574.22 & Fe I &  3.21 & -2.50 &   36.5 &    7.50 &      \\
 4602.01 & Fe I &  1.61 & -3.15 &   72.4 &    7.49 &      \\
 4785.96 & Fe I &  4.14 & -1.93 &   26.1 &    7.55 &      \\
 4808.15 & Fe I &  3.25 & -2.79 &   26.4 &    7.57 &      \\
 4874.36 & Fe I &  3.07 & -3.03 &   21.7 &    7.52 &      \\
 4892.86 & Fe I &  4.22 & -1.29 &   52.8 &    7.51 &      \\
 4962.58 & Fe I &  4.18 & -1.18 &   55.8 &    7.43 &      \\
 4992.79 & Fe I &  4.26 & -2.35 &   10.9 &    7.58 &      \\
 5016.48 & Fe I &  4.26 & -1.69 &   30.6 &    7.50 &      \\
 5044.22 & Fe I &  2.85 & -2.04 &   76.2 &    7.54 &      \\
 5058.50 & Fe I &  3.64 & -2.83 &   11.3 &    7.49 &      \\
 5247.06 & Fe I &  0.09 & -4.95 &   67.6 &    7.56 &      \\
 5250.22 & Fe I &  0.12 & -4.94 &   65.6 &    7.53 &      \\
 5253.02 & Fe I &  2.28 & -3.94 &   19.3 &    7.57 &      \\
 5295.32 & Fe I &  4.41 & -1.69 &   28.7 &    7.60 &      \\
 5320.04 & Fe I &  3.64 & -2.54 &   18.9 &    7.47 &      \\
 5379.58 & Fe I &  3.69 & -1.51 &   61.4 &    7.44 &      \\
 5386.34 & Fe I &  4.15 & -1.77 &   31.7 &    7.50 &      \\
 5522.45 & Fe I &  4.21 & -1.55 &   45.7 &    7.62 &      \\
 5539.28 & Fe I &  3.64 & -2.66 &   19.1 &    7.58 &      \\
 5560.22 & Fe I &  4.43 & -1.19 &   53.2 &    7.60 &      \\
 5577.02 & Fe I &  5.03 & -1.55 &   10.7 &    7.47 &      \\
 5579.34 & Fe I &  4.23 & -2.40 &    8.8 &    7.48 &      \\
 5636.70 & Fe I &  3.64 & -2.61 &   18.8 &    7.52 &      \\
 5709.93 & Fe I &  4.26 & -2.34 &   11.7 &    7.58 &      \\
 5811.92 & Fe I &  4.14 & -2.43 &   11.8 &    7.56 &      \\
 5814.81 & Fe I &  4.28 & -1.97 &   22.4 &    7.58 &      \\
 5927.79 & Fe I &  4.65 & -1.09 &   41.5 &    7.46 &      \\
 5934.66 & Fe I &  3.93 & -1.17 &   74.4 &    7.51 &      \\
 5956.70 & Fe I &  0.86 & -4.61 &   54.0 &    7.57 &      \\
 6008.57 & Fe I &  3.88 & -0.99 &   89.6 &    7.53 &      \\
 6079.01 & Fe I &  4.65 & -1.12 &   46.3 &    7.58 &      \\
 6082.72 & Fe I &  2.22 & -3.57 &   35.1 &    7.49 &      \\
 6093.65 & Fe I &  4.61 & -1.50 &   30.0 &    7.59 &      \\
 6098.25 & Fe I &  4.56 & -1.88 &   16.0 &    7.55 &      \\
 6151.62 & Fe I &  2.18 & -3.30 &   49.5 &    7.48 &      \\
 6173.34 & Fe I &  2.22 & -2.88 &   69.9 &    7.54 &      \\
 6187.99 & Fe I &  3.94 & -1.72 &   47.7 &    7.56 &      \\
 6200.32 & Fe I &  2.61 & -2.44 &   74.0 &    7.56 &      \\
 6220.78 & Fe I &  3.88 & -2.46 &   17.8 &    7.54 &      \\
 6226.74 & Fe I &  3.88 & -2.22 &   29.2 &    7.61 &      \\
 6265.14 & Fe I &  2.18 & -2.55 &   85.9 &    7.49 &      \\
 6311.50 & Fe I &  2.83 & -3.14 &   28.9 &    7.50 &      \\
 6322.69 & Fe I &  2.59 & -2.43 &   77.6 &    7.59 &      \\
 6344.16 & Fe I &  2.43 & -2.92 &   61.5 &    7.60 &      \\
 6481.88 & Fe I &  2.28 & -2.98 &   66.7 &    7.61 &      \\
 6574.23 & Fe I &  0.99 & -5.02 &   26.7 &    7.48 &      \\
 6581.21 & Fe I &  1.49 & -4.68 &   19.9 &    7.45 &      \\
 6591.31 & Fe I &  4.59 & -2.07 &   10.1 &    7.52 &      \\
 6593.88 & Fe I &  2.43 & -2.42 &   84.6 &    7.54 &      \\
 6608.03 & Fe I &  2.28 & -4.03 &   16.2 &    7.48 &      \\
 6710.32 & Fe I &  1.49 & -4.88 &   17.0 &    7.56 &      \\
 6725.36 & Fe I &  4.10 & -2.30 &   17.1 &    7.55 &      \\
 6733.15 & Fe I &  4.64 & -1.58 &   26.5 &    7.59 &      \\
 6737.99 & Fe I &  4.56 & -1.75 &   21.5 &    7.56 &      \\
 6750.16 & Fe I &  2.42 & -2.62 &   75.9 &    7.55 &      \\
 6806.85 & Fe I &  2.73 & -3.21 &   34.6 &    7.57 &      \\
 6810.27 & Fe I &  4.61 & -0.99 &   50.8 &    7.46 &      \\
 6839.84 & Fe I &  2.56 & -3.45 &   29.3 &    7.52 &      \\
 6842.69 & Fe I &  4.64 & -1.32 &   37.1 &    7.56 &      \\
 6855.17 & Fe I &  4.56 & -0.74 &   70.8 &    7.51 &      \\
 6855.72 & Fe I &  4.61 & -1.82 &   16.7 &    7.53 &      \\
 6857.25 & Fe I &  4.08 & -2.15 &   23.1 &    7.54 &      \\
 6862.50 & Fe I &  4.56 & -1.57 &   29.6 &    7.58 &      \\
 7022.96 & Fe I &  4.19 & -1.25 &   58.9 &    7.49 &      \\
 7024.07 & Fe I &  4.08 & -2.21 &   20.9 &    7.54 &      \\
 7219.69 & Fe I &  4.08 & -1.69 &   47.4 &    7.59 &      \\
 7256.14 & Fe I &  4.96 & -1.59 &   17.3 &    7.63 &      \\
 7421.56 & Fe I &  4.64 & -1.80 &   14.5 &    7.44 &      \\
 7430.54 & Fe I &  2.59 & -3.86 &   13.2 &    7.46 &      \\
 7461.53 & Fe I &  2.56 & -3.58 &   26.2 &    7.54 &      \\
 7498.53 & Fe I &  4.14 & -2.25 &   18.0 &    7.53 &      \\
 7540.44 & Fe I &  2.73 & -3.85 &   10.7 &    7.48 &      \\
 7547.90 & Fe I &  5.10 & -1.35 &   20.0 &    7.60 &      \\
 7723.21 & Fe I &  2.28 & -3.62 &   42.5 &    7.64 &      \\
 7745.52 & Fe I &  5.09 & -1.17 &   23.5 &    7.49 &      \\
 7746.60 & Fe I &  5.06 & -1.28 &   18.9 &    7.45 &      \\
 7751.11 & Fe I &  4.99 & -0.75 &   48.4 &    7.48 &      \\
 7807.91 & Fe I &  4.99 & -0.54 &   61.1 &    7.48 &      \\
 7844.56 & Fe I &  4.84 & -1.81 &   12.3 &    7.54 &      \\
 7912.87 & Fe I &  0.86 & -4.85 &   47.0 &    7.51 &      \\
\multicolumn{7}{l}{\textbf{Fe II  A$_\odot$ =  7.53}}  \\
 5132.67 &Fe II &  2.81 & -4.09 &   24.5 &    7.53 &      \\
 5325.56 &Fe II &  3.22 & -3.32 &   39.9 &    7.55 &      \\
 5414.07 &Fe II &  3.22 & -3.64 &   25.8 &    7.50 &      \\
 6084.11 &Fe II &  3.20 & -3.88 &   20.1 &    7.53 &      \\
 6149.25 &Fe II &  3.89 & -2.84 &   34.9 &    7.55 &      \\
 6369.46 &Fe II &  2.89 & -4.23 &   18.0 &    7.52 &      \\
 6456.39 &Fe II &  3.90 & -2.18 &   60.8 &    7.52 &      \\
\multicolumn{7}{l}{\textbf{Co I  A$_\odot$ =  4.88}}   \\
 5342.70 & Co I &  4.02 &  0.69 &   32.0 &    4.82 &      \\
 5647.23 & Co I &  2.28 & -1.56 &   12.7 &    4.85 &      \\
 6093.15 & Co I &  1.74 & -2.44 &    7.7 &    4.93 &      \\
 6455.00 & Co I &  3.63 & -0.25 &   12.6 &    4.80 &      \\
 6632.45 & Co I &  2.28 & -2.00 &    7.1 &    4.95 &      \\
\multicolumn{7}{l}{\textbf{Ni I  A$_\odot$ =  6.28}}   \\
 5578.72 & Ni I &  1.68 & -2.64 &   54.2 &    6.21 &      \\
 5589.36 & Ni I &  3.90 & -1.14 &   27.8 &    6.26 &      \\
 5593.74 & Ni I &  3.90 & -0.84 &   45.8 &    6.33 &      \\
 5625.32 & Ni I &  4.09 & -0.70 &   38.6 &    6.22 &      \\
 5682.20 & Ni I &  4.10 & -0.47 &   53.5 &    6.28 &      \\
 5694.99 & Ni I &  4.09 & -0.61 &   47.7 &    6.30 &      \\
 5748.35 & Ni I &  1.68 & -3.26 &   26.9 &    6.22 &      \\
 5760.83 & Ni I &  4.10 & -0.81 &   35.9 &    6.28 &      \\
 5796.09 & Ni I &  1.95 & -3.69 &    8.1 &    6.26 &      \\
 5805.22 & Ni I &  4.17 & -0.64 &   44.1 &    6.33 &      \\
 6053.69 & Ni I &  4.23 & -1.07 &   20.1 &    6.28 &      \\
 6086.28 & Ni I &  4.26 & -0.51 &   47.3 &    6.35 &      \\
 6111.07 & Ni I &  4.09 & -0.87 &   35.6 &    6.31 &      \\
 6128.97 & Ni I &  1.68 & -3.33 &   24.4 &    6.21 &      \\
 6130.14 & Ni I &  4.27 & -0.96 &   22.5 &    6.27 &      \\
 6175.37 & Ni I &  4.09 & -0.54 &   53.8 &    6.33 &      \\
 6177.24 & Ni I &  1.83 & -3.51 &   13.6 &    6.20 &      \\
 6186.71 & Ni I &  4.10 & -0.96 &   32.7 &    6.36 &      \\
 6204.60 & Ni I &  4.09 & -1.14 &   22.2 &    6.27 &      \\
 6370.35 & Ni I &  3.54 & -1.94 &   11.5 &    6.19 &      \\
 6378.25 & Ni I &  4.15 & -0.90 &   32.7 &    6.33 &      \\
 6586.31 & Ni I &  1.95 & -2.81 &   41.4 &    6.30 &      \\
 6598.60 & Ni I &  4.23 & -0.98 &   25.0 &    6.30 &      \\
 6635.12 & Ni I &  4.42 & -0.83 &   24.7 &    6.32 &      \\
 6767.77 & Ni I &  1.83 & -2.17 &   82.3 &    6.38 &      \\
 6772.31 & Ni I &  3.66 & -0.99 &   51.9 &    6.33 &      \\
 6842.04 & Ni I &  3.66 & -1.47 &   24.0 &    6.22 &      \\
 7030.01 & Ni I &  3.54 & -1.73 &   21.2 &    6.28 &      \\
 7110.88 & Ni I &  1.93 & -2.97 &   36.7 &    6.32 &      \\
 7422.27 & Ni I &  3.63 & -0.13 &  102.6 &    6.25 &      \\
 7574.05 & Ni I &  3.83 & -0.58 &   68.0 &    6.33 &      \\
 7727.62 & Ni I &  3.68 & -0.16 &  100.8 &    6.27 &      \\
 7748.89 & Ni I &  3.70 & -0.13 &   96.0 &    6.19 &      \\
 7797.59 & Ni I &  3.90 & -0.18 &   87.2 &    6.28 &      \\
 7826.77 & Ni I &  3.70 & -1.95 &   12.1 &    6.31 &      \\

\enddata
\end{deluxetable} 

%% file: table5.tex
\begin{deluxetable}{lr@{}l@{~~}r@{}l@{~~}r@{}l@{~~}r@{}l@{~~}r@{}l@{~~}r@{}l@{~~}r@{}l@{~~}r@{}l} 
\tabletypesize{\scriptsize}
\tablecaption{Chemical Abundances of Na, Mg, Al, Si, Ca, and Sc}
\tablewidth{0cm}
\tablehead{
  \multicolumn{1}{c}{Star} &
  \multicolumn{2}{c}{[Na/H]} &
  \multicolumn{2}{c}{[Mg/H]} &
  \multicolumn{2}{c}{[Al/H]} &
  \multicolumn{2}{c}{[Si/H]} &
  \multicolumn{2}{c}{[Ca/H]} &
  \multicolumn{2}{c}{[Sc/H]} &
}
\startdata
\multicolumn{13}{l}{\textbf{34 Planet Host Stars}} \\
      HD 10697 &   0.18&$\pm$0.04 &   0.13&$\pm$0.03 &   0.19&$\pm$0.06 &   0.14&$\pm$0.05 &   0.12&$\pm$0.08 &   0.27&$\pm$0.04  \\
      HD 16141 &   0.05&$\pm$0.12 &   0.14&$\pm$0.04 &   0.09&$\pm$0.03 &   0.07&$\pm$0.04 &   0.11&$\pm$0.05 &   0.21&$\pm$0.05  \\
      HD 16400 &   0.16&          &   0.08&$\pm$0.01 &   0.10&$\pm$0.05 &   0.11&$\pm$0.08 &   0.07&$\pm$0.04 &   0.12&$\pm$0.09  \\
      HD 17156 &   0.27&$\pm$0.03 &   0.23&$\pm$0.16 &   0.29&$\pm$0.11 &   0.23&$\pm$0.07 &   0.21&$\pm$0.12 &   0.33&$\pm$0.10  \\
     BD+20 518 &   0.36&$\pm$0.03 &   0.26&$\pm$0.03 &   0.28&$\pm$0.06 &   0.35&$\pm$0.10 &   0.31&$\pm$0.11 &   0.54&$\pm$0.00  \\
      HD 20367 &   0.05&$\pm$0.03 &   0.02&          &  -0.01&$\pm$0.03 &   0.07&$\pm$0.10 &   0.17&$\pm$0.10 &   0.09&$\pm$0.03  \\
      HD 28305 &   0.33&$\pm$0.06 &   0.14&$\pm$0.12 &   0.28&$\pm$0.06 &   0.17&$\pm$0.03 &   0.06&$\pm$0.05 &   0.17&$\pm$0.12  \\
      HD 37124 &  -0.40&$\pm$0.05 &  -0.11&$\pm$0.13 &  -0.18&$\pm$0.06 &  -0.33&$\pm$0.06 &  -0.35&$\pm$0.04 &  -0.19&$\pm$0.01  \\
      HD 38529 &   0.50&$\pm$0.17 &   0.39&$\pm$0.12 &   0.45&$\pm$0.12 &   0.37&$\pm$0.12 &   0.32&$\pm$0.11 &   0.64&$\pm$0.21  \\
      HD 43691 &   0.37&$\pm$0.03 &   0.17&$\pm$0.11 &   0.25&$\pm$0.03 &   0.25&$\pm$0.08 &   0.27&$\pm$0.10 &   0.38&$\pm$0.06  \\
      HD 45350 &   0.24&$\pm$0.03 &   0.29&$\pm$0.05 &   0.30&$\pm$0.04 &   0.23&$\pm$0.07 &   0.24&$\pm$0.09 &   0.42&$\pm$0.09  \\
      HD 52265 &   0.25&$\pm$0.03 &   0.18&$\pm$0.10 &   0.17&$\pm$0.06 &   0.25&$\pm$0.10 &   0.23&$\pm$0.09 &   0.27&$\pm$0.04  \\
      HD 74156 &   0.17&$\pm$0.04 &   0.11&$\pm$0.05 &   0.10&$\pm$0.01 &   0.09&$\pm$0.09 &   0.11&$\pm$0.10 &   0.22&$\pm$0.05  \\
      HD 75732 &   0.37&$\pm$0.03 &   0.35&$\pm$0.04 &   0.44&$\pm$0.08 &   0.42&$\pm$0.10 &   0.30&$\pm$0.09 &   0.35&$\pm$0.02  \\
      HD 75898 &   0.33&$\pm$0.05 &   0.15&          &   0.22&$\pm$0.02 &   0.25&$\pm$0.10 &   0.30&$\pm$0.11 &   0.38&$\pm$0.10  \\
      HD 81040 &  -0.24&$\pm$0.04 &  -0.05&$\pm$0.03 &  -0.19&$\pm$0.02 &  -0.12&$\pm$0.09 &  -0.01&$\pm$0.05 &  -0.05&$\pm$0.04  \\
      HD 88133 &   0.26&$\pm$0.09 &   0.36&$\pm$0.10 &   0.39&$\pm$0.05 &   0.34&$\pm$0.12 &   0.31&$\pm$0.12 &   0.52&$\pm$0.08  \\
      HD 89307 &  -0.16&$\pm$0.05 &  -0.08&$\pm$0.11 &  -0.16&$\pm$0.03 &  -0.12&$\pm$0.06 &  -0.11&$\pm$0.10 &  -0.04&$\pm$0.02  \\
      HD 92788 &   0.27&$\pm$0.07 &   0.24&$\pm$0.12 &   0.30&$\pm$0.04 &   0.27&$\pm$0.10 &   0.23&$\pm$0.10 &   0.45&$\pm$0.01  \\
      HD 95128 &   0.01&$\pm$0.06 &   0.01&$\pm$0.07 &   0.00&$\pm$0.01 &   0.02&$\pm$0.09 &  -0.06&$\pm$0.10 &   0.07&$\pm$0.06  \\
     HD 104985 &  -0.23&$\pm$0.03 &  -0.04&$\pm$0.03 &   0.03&$\pm$0.02 &  -0.00&$\pm$0.12 &  -0.26&$\pm$0.02 &   0.04&$\pm$0.17  \\
     HD 106252 &  -0.04&$\pm$0.04 &  -0.16&$\pm$0.06 &   0.01&$\pm$0.08 &  -0.11&$\pm$0.09 &  -0.07&$\pm$0.08 &   0.05&$\pm$0.09  \\
     HD 117176 &  -0.18&$\pm$0.06 &   0.01&$\pm$0.07 &   0.02&$\pm$0.04 &  -0.05&$\pm$0.09 &  -0.06&$\pm$0.11 &   0.15&$\pm$0.02  \\
     HD 143761 &  -0.21&$\pm$0.02 &  -0.01&$\pm$0.18 &  -0.05&$\pm$0.06 &  -0.16&$\pm$0.09 &  -0.12&$\pm$0.08 &  -0.02&$\pm$0.06  \\
     HD 149026 &   0.42&$\pm$0.04 &   0.13&          &   0.28&$\pm$0.07 &   0.33&$\pm$0.09 &   0.31&$\pm$0.08 &   0.55&$\pm$0.03  \\
     HD 154345 &  -0.24&$\pm$0.06 &  -0.11&$\pm$0.07 &  -0.13&$\pm$0.04 &  -0.15&$\pm$0.09 &  -0.12&$\pm$0.10 &  -0.02&$\pm$0.08  \\
     HD 155358 &  -0.58&$\pm$0.04 &  -0.47&$\pm$0.08 &  -0.55&$\pm$0.07 &  -0.53&$\pm$0.09 &  -0.47&$\pm$0.09 &  -0.36&$\pm$0.08  \\
     HD 185269 &   0.14&$\pm$0.01 &   0.04&$\pm$0.04 &   0.10&$\pm$0.03 &   0.09&$\pm$0.07 &   0.19&$\pm$0.10 &   0.25&$\pm$0.05  \\
     HD 186427 &   0.03&$\pm$0.03 &   0.04&$\pm$0.06 &   0.06&$\pm$0.02 &   0.03&$\pm$0.05 &   0.02&$\pm$0.10 &   0.20&$\pm$0.08  \\
     HD 190228 &  -0.32&$\pm$0.05 &  -0.16&$\pm$0.07 &  -0.23&$\pm$0.04 &  -0.29&$\pm$0.10 &  -0.22&$\pm$0.10 &  -0.16&$\pm$0.06  \\
     HD 190360 &   0.19&$\pm$0.06 &   0.27&$\pm$0.05 &   0.31&$\pm$0.05 &   0.27&$\pm$0.08 &   0.19&$\pm$0.11 &   0.38&$\pm$0.04  \\
     HD 195019 &  -0.01&$\pm$0.04 &   0.03&$\pm$0.07 &   0.05&$\pm$0.02 &   0.01&$\pm$0.08 &   0.03&$\pm$0.11 &   0.17&$\pm$0.11  \\
     HD 209458 &  -0.04&$\pm$0.04 &  -0.09&          &  -0.03&$\pm$0.06 &  -0.01&$\pm$0.08 &   0.00&$\pm$0.07 &   0.08&$\pm$0.06  \\
     HD 217014 &   0.21&$\pm$0.05 &   0.17&$\pm$0.07 &   0.22&$\pm$0.02 &   0.15&$\pm$0.05 &   0.19&$\pm$0.07 &   0.33&$\pm$0.07  \\

\multicolumn{13}{l}{\textbf{18 Comparison Stars}} \\

      HD 10307 &   0.04&$\pm$0.05 &  -0.00&$\pm$0.09 &   0.03&$\pm$0.02 &   0.03&$\pm$0.07 &   0.02&$\pm$0.11 &   0.09&$\pm$0.08  \\
      HD 13974 &  -0.34&$\pm$0.16 &  -0.57&          &  -0.55&$\pm$0.02 &  -0.49&$\pm$0.11 &  -0.40&$\pm$0.12 &  -0.54&$\pm$0.06  \\
      HD 24040 &   0.18&$\pm$0.05 &   0.20&$\pm$0.08 &   0.21&$\pm$0.08 &   0.16&$\pm$0.07 &   0.18&$\pm$0.10 &   0.37&$\pm$0.07  \\
      HD 26722 &   0.11&$\pm$0.12 &  -0.11&$\pm$0.05 &  -0.10&$\pm$0.06 &  -0.00&$\pm$0.11 &  -0.10&$\pm$0.10 &   0.05&$\pm$0.18  \\
      HD 32923 &  -0.20&$\pm$0.05 &  -0.09&$\pm$0.04 &  -0.03&$\pm$0.02 &  -0.12&$\pm$0.10 &  -0.13&$\pm$0.09 &  -0.04&$\pm$0.04  \\
      HD 33636 &  -0.17&$\pm$0.04 &  -0.03&$\pm$0.13 &  -0.16&$\pm$0.02 &  -0.13&$\pm$0.07 &  -0.13&$\pm$0.08 &   0.01&$\pm$0.00  \\
      HD 39587 &  -0.16&$\pm$0.02 &  -0.11&$\pm$0.10 &  -0.11&          &  -0.10&$\pm$0.07 &   0.01&$\pm$0.07 &   0.05&$\pm$0.04  \\
      HD 48682 &   0.09&$\pm$0.06 &   0.03&$\pm$0.03 &   0.01&$\pm$0.03 &   0.08&$\pm$0.09 &   0.11&$\pm$0.08 &   0.12&$\pm$0.05  \\
      HD 50692 &  -0.16&$\pm$0.03 &  -0.03&$\pm$0.10 &  -0.13&$\pm$0.07 &  -0.13&$\pm$0.09 &  -0.08&$\pm$0.10 &   0.01&$\pm$0.03  \\
      HD 55575 &  -0.36&$\pm$0.03 &  -0.04&$\pm$0.08 &  -0.29&$\pm$0.07 &  -0.30&$\pm$0.07 &  -0.22&$\pm$0.07 &  -0.20&$\pm$0.06  \\
      HD 72905 &  -0.17&$\pm$0.04 &  -0.12&$\pm$0.16 &  -0.11&$\pm$0.01 &  -0.13&$\pm$0.07 &  -0.04&$\pm$0.06 &  -0.06&$\pm$0.09  \\
      HD 84737 &   0.10&$\pm$0.06 &   0.15&$\pm$0.09 &   0.12&$\pm$0.04 &   0.13&$\pm$0.10 &   0.14&$\pm$0.10 &   0.28&$\pm$0.09  \\
     HD 109358 &  -0.18&$\pm$0.04 &  -0.13&$\pm$0.06 &  -0.21&$\pm$0.01 &  -0.22&$\pm$0.09 &  -0.21&$\pm$0.11 &  -0.14&$\pm$0.01  \\
     HD 110897 &  -0.61&$\pm$0.06 &  -0.42&$\pm$0.06 &  -0.53&$\pm$0.07 &  -0.48&$\pm$0.05 &  -0.47&$\pm$0.08 &  -0.53&$\pm$0.03  \\
     HD 137510 &   0.50&$\pm$0.10 &   0.16&$\pm$0.08 &   0.32&$\pm$0.04 &   0.33&$\pm$0.08 &   0.25&$\pm$0.10 &   0.42&$\pm$0.02  \\
     HD 141004 &  -0.01&$\pm$0.05 &   0.03&$\pm$0.05 &   0.03&$\pm$0.01 &  -0.02&$\pm$0.09 &   0.02&$\pm$0.09 &   0.14&$\pm$0.05  \\*
     HD 161797 &   0.34&$\pm$0.13 &   0.26&$\pm$0.01 &   0.36&$\pm$0.02 &   0.30&$\pm$0.10 &   0.27&$\pm$0.09 &   0.47&$\pm$0.10  \\*
     HD 188512 &  -0.27&$\pm$0.11 &  -0.08&$\pm$0.07 &  -0.13&$\pm$0.04 &  -0.16&$\pm$0.10 &  -0.13&$\pm$0.11 &   0.02&$\pm$0.08  \\*
\enddata
\end{deluxetable}

%% file: table6.tex
\begin{deluxetable}{lr@{}l@{~~}r@{}l@{~~}r@{}l@{~~}r@{}l@{~~}r@{}l@{~~}r@{}l@{~~}r@{}l} 
\tabletypesize{\scriptsize}
\tablecaption{Chemical abundances for Ti, V, Cr, Mn, Co, and Ni}
\tablewidth{0cm}
\tablehead{
  \multicolumn{1}{c}{Star} &
  \multicolumn{2}{c}{[Ti$_\mathrm{I}$/H]} &   
  \multicolumn{2}{c}{[Ti$_\mathrm{II}$/H]} &    
  \multicolumn{2}{c}{[V/H]} &    
  \multicolumn{2}{c}{[Cr/H]} &
  \multicolumn{2}{c}{[Mn/H]} &
  \multicolumn{2}{c}{[Co/H]} &
  \multicolumn{2}{c}{[Ni/H]}
}
\startdata
\multicolumn{15}{l}{\textbf{34 Planet Host Stars}} \\
      HD 10697&   0.19&$\pm$0.07 &   0.20&$\pm$0.10 &   0.16&$\pm$0.05 &   0.08&$\pm$0.05 &   0.71&          &   0.19&$\pm$0.06 &   0.18&$\pm$0.07 \\ 
      HD 16141&   0.17&$\pm$0.06 &   0.22&$\pm$0.07 &   0.13&$\pm$0.06 &   0.07&$\pm$0.05 &   0.04&$\pm$0.03 &   0.14&$\pm$0.07 &   0.12&$\pm$0.05 \\ 
      HD 16400&   0.17&$\pm$0.07 &   0.14&$\pm$0.03 &   0.27&$\pm$0.09 &   0.13&$\pm$0.03 &   0.05&          &   0.29&$\pm$0.11 &   0.08&$\pm$0.07 \\ 
      HD 17156&   0.29&$\pm$0.10 &   0.31&$\pm$0.04 &   0.29&$\pm$0.04 &   0.22&$\pm$0.08 &   0.49&          &   0.20&$\pm$0.09 &   0.25&$\pm$0.07 \\ 
     BD+20 518&   0.35&$\pm$0.11 &   0.18&$\pm$0.08 &   0.43&$\pm$0.09 &   0.30&$\pm$0.11 &   0.32&          &   0.47&$\pm$0.08 &   0.40&$\pm$0.06 \\ 
      HD 20367&   0.18&$\pm$0.09 &   0.15&$\pm$0.13 &   0.18&$\pm$0.05 &   0.14&$\pm$0.08 &   0.18&$\pm$0.12 &   0.06&$\pm$0.09 &   0.07&$\pm$0.06 \\ 
      HD 28305&   0.20&$\pm$0.08 &   0.10&$\pm$0.17 &   0.39&$\pm$0.17 &   0.18&$\pm$0.11 &   0.22&          &   0.41&$\pm$0.09 &   0.14&$\pm$0.05 \\ 
      HD 37124&  -0.12&$\pm$0.05 &  -0.19&$\pm$0.08 &  -0.17&$\pm$0.07 &  -0.49&$\pm$0.01 &  -0.44&$\pm$0.06 &  -0.28&$\pm$0.03 &  -0.42&$\pm$0.05 \\ 
      HD 38529&   0.44&$\pm$0.12 &   0.40&$\pm$0.18 &   0.48&$\pm$0.06 &   0.30&$\pm$0.03 &   0.36&$\pm$0.05 &   0.50&$\pm$0.08 &   0.45&$\pm$0.06 \\ 
      HD 43691&   0.28&$\pm$0.05 &   0.35&$\pm$0.07 &   0.33&$\pm$0.05 &   0.26&$\pm$0.09 &   0.18&$\pm$0.06 &   0.29&$\pm$0.10 &   0.32&$\pm$0.05 \\ 
      HD 45350&   0.36&$\pm$0.06 &   0.34&$\pm$0.12 &   0.37&$\pm$0.03 &   0.23&$\pm$0.01 &   0.26&$\pm$0.05 &   0.36&$\pm$0.09 &   0.32&$\pm$0.04 \\ 
      HD 52265&   0.30&$\pm$0.09 &   0.27&$\pm$0.15 &   0.28&$\pm$0.05 &   0.20&$\pm$0.09 &   0.38&          &   0.16&$\pm$0.02 &   0.24&$\pm$0.06 \\ 
      HD 74156&   0.22&$\pm$0.06 &   0.20&$\pm$0.11 &   0.17&$\pm$0.04 &   0.11&$\pm$0.10 &   0.35&          &   0.09&$\pm$0.10 &   0.15&$\pm$0.05 \\ 
      HD 75732&   0.38&$\pm$0.05 &   0.16&$\pm$0.12 &   0.65&$\pm$0.18 &   0.37&$\pm$0.06 &   0.37&          &   0.53&$\pm$0.11 &   0.43&$\pm$0.08 \\ 
      HD 75898&   0.32&$\pm$0.10 &   0.43&$\pm$0.08 &   0.37&$\pm$0.05 &   0.25&$\pm$0.09 &   0.56&          &   0.33&$\pm$0.06 &   0.32&$\pm$0.05 \\ 
      HD 81040&  -0.02&$\pm$0.08 &  -0.04&$\pm$0.09 &  -0.06&$\pm$0.09 &  -0.12&$\pm$0.06 &   0.13&          &  -0.17&$\pm$0.09 &  -0.16&$\pm$0.08 \\ 
      HD 88133&   0.36&$\pm$0.12 &   0.43&$\pm$0.14 &   0.43&$\pm$0.07 &   0.30&$\pm$0.05 &   0.34&$\pm$0.05 &   0.54&$\pm$0.10 &   0.41&$\pm$0.07 \\ 
      HD 89307&   0.01&$\pm$0.08 &   0.01&$\pm$0.05 &  -0.15&$\pm$0.05 &  -0.13&$\pm$0.05 &  -0.07&$\pm$0.07 &  -0.18&$\pm$0.09 &  -0.17&$\pm$0.04 \\ 
      HD 92788&   0.31&$\pm$0.10 &   0.29&$\pm$0.12 &   0.34&$\pm$0.03 &   0.20&$\pm$0.03 &   0.26&          &   0.38&$\pm$0.12 &   0.34&$\pm$0.08 \\ 
      HD 95128&   0.07&$\pm$0.09 &   0.04&$\pm$0.09 &   0.02&$\pm$0.06 &  -0.03&$\pm$0.07 &   0.06&$\pm$0.13 &  -0.01&$\pm$0.08 &  -0.02&$\pm$0.07 \\ 
     HD 104985&  -0.03&$\pm$0.06 &  -0.06&$\pm$0.13 &   0.12&$\pm$0.17 &  -0.34&$\pm$0.12 &  -0.43&          &   0.06&$\pm$0.12 &  -0.25&$\pm$0.06 \\ 
     HD 106252&   0.05&$\pm$0.08 &   0.09&$\pm$0.02 &   0.04&$\pm$0.08 &  -0.06&$\pm$0.10 &   0.15&          &  -0.02&$\pm$0.09 &  -0.05&$\pm$0.04 \\ 
     HD 117176&   0.04&$\pm$0.08 &   0.13&$\pm$0.09 &   0.04&$\pm$0.08 &  -0.07&$\pm$0.09 &  -0.11&          &   0.01&$\pm$0.09 &  -0.08&$\pm$0.05 \\ 
     HD 143761&   0.05&$\pm$0.07 &   0.02&$\pm$0.09 &  -0.05&$\pm$0.06 &  -0.20&$\pm$0.07 &  -0.14&          &  -0.10&$\pm$0.12 &  -0.20&$\pm$0.03 \\ 
     HD 149026&   0.47&$\pm$0.08 &   0.56&$\pm$0.06 &   0.43&$\pm$0.04 &   0.36&$\pm$0.12 &   0.69&          &   0.41&$\pm$0.06 &   0.44&$\pm$0.06 \\ 
     HD 154345&   0.04&$\pm$0.11 &  -0.05&$\pm$0.10 &   0.01&$\pm$0.08 &  -0.07&$\pm$0.08 &  -0.06&          &  -0.02&$\pm$0.05 &  -0.10&$\pm$0.05 \\ 
     HD 155358&  -0.36&$\pm$0.09 &  -0.28&$\pm$0.12 &  -0.47&$\pm$0.16 &  -0.58&$\pm$0.14 &  -0.81&$\pm$0.11 &  -0.62&$\pm$0.06 &  -0.69&$\pm$0.03 \\ 
     HD 185269&   0.19&$\pm$0.08 &   0.34&$\pm$0.12 &   0.17&$\pm$0.10 &   0.10&$\pm$0.10 &   0.26&          &   0.13&$\pm$0.02 &   0.13&$\pm$0.02 \\ 
     HD 186427&   0.14&$\pm$0.05 &   0.15&$\pm$0.10 &   0.12&$\pm$0.08 &   0.05&$\pm$0.09 &   0.53&          &   0.15&$\pm$0.08 &   0.08&$\pm$0.04 \\ 
     HD 190228&  -0.19&$\pm$0.05 &  -0.09&$\pm$0.17 &  -0.21&$\pm$0.02 &  -0.37&$\pm$0.04 &  -0.29&          &  -0.23&$\pm$0.06 &  -0.33&$\pm$0.03 \\ 
     HD 190360&   0.32&$\pm$0.11 &   0.30&$\pm$0.11 &   0.35&$\pm$0.05 &   0.17&$\pm$0.04 &   \multicolumn{2}{c}{\nodata} &   0.40&$\pm$0.11 &   0.28&$\pm$0.01 \\
     HD 195019&   0.10&$\pm$0.07 &   0.06&$\pm$0.08 &   0.06&$\pm$0.04 &  -0.01&$\pm$0.06 &   0.26&          &   0.01&$\pm$0.06 &   0.01&$\pm$0.04 \\ 
     HD 209458&   0.12&$\pm$0.10 &   0.07&$\pm$0.09 &   0.03&$\pm$0.05 &   0.02&$\pm$0.05 &   0.07&          &  -0.02&$\pm$0.10 &  -0.02&$\pm$0.04 \\ 
     HD 217014&   0.30&$\pm$0.08 &   0.24&$\pm$0.09 &   0.29&$\pm$0.05 &   0.15&$\pm$0.01 &   0.22&$\pm$0.05 &   0.30&$\pm$0.07 &   0.28&$\pm$0.07 \\ 
                                                                                                                                                      
\multicolumn{15}{l}{\textbf{18 Comparison Stars}} \\                                                                                                                                       
                                                                                                                                                      
      HD 10307&   0.08&$\pm$0.10 &   0.12&$\pm$0.05 &   0.11&$\pm$0.05 &   0.02&$\pm$0.08 &  -0.01&$\pm$0.03 &   0.05&$\pm$0.04 &   0.04&$\pm$0.04 \\ 
      HD 13974&  -0.38&$\pm$0.08 &  -0.30&$\pm$0.13 &  -0.48&$\pm$0.05 &  -0.54&$\pm$0.06 &  -0.51&          &  -0.56&$\pm$0.08 &  -0.53&$\pm$0.03 \\ 
      HD 24040&   0.29&$\pm$0.06 &   0.39&$\pm$0.07 &   0.34&$\pm$0.08 &   0.21&$\pm$0.08 &   0.07&$\pm$0.06 &   0.34&$\pm$0.04 &   0.26&$\pm$0.06 \\ 
      HD 26722&  -0.04&$\pm$0.12 &  -0.01&$\pm$0.18 &  -0.12&$\pm$0.05 &  -0.19&$\pm$0.06 &  -0.39&$\pm$0.12 &  -0.04&$\pm$0.15 &  -0.15&$\pm$0.06 \\ 
      HD 32923&   0.01&$\pm$0.08 &   0.02&$\pm$0.05 &  -0.12&$\pm$0.08 &  -0.23&$\pm$0.06 &  -0.25&          &  -0.14&$\pm$0.09 &  -0.21&$\pm$0.04 \\ 
      HD 33636&   0.02&$\pm$0.09 &  -0.01&$\pm$0.08 &  -0.07&$\pm$0.06 &  -0.14&$\pm$0.06 &  -0.25&          &  -0.09&$\pm$0.10 &  -0.15&$\pm$0.04 \\ 
      HD 39587&   0.10&$\pm$0.09 &   0.15&$\pm$0.08 &   0.03&$\pm$0.05 &   0.04&$\pm$0.08 &  -0.12&          &  -0.03&          &  -0.08&$\pm$0.05 \\ 
      HD 48682&   0.16&$\pm$0.10 &   0.20&$\pm$0.08 &   0.09&$\pm$0.03 &   0.01&$\pm$0.02 &   0.03&          &   0.07&$\pm$0.02 &   0.09&$\pm$0.05 \\ 
      HD 50692&  -0.02&$\pm$0.10 &  -0.07&$\pm$0.09 &  -0.13&$\pm$0.10 &  -0.16&$\pm$0.07 &  -0.23&          &  -0.11&$\pm$0.10 &  -0.17&$\pm$0.05 \\ 
      HD 55575&  -0.14&$\pm$0.08 &  -0.10&$\pm$0.08 &  -0.28&$\pm$0.06 &  -0.39&$\pm$0.06 &  -0.55&          &  -0.32&$\pm$0.05 &  -0.38&$\pm$0.06 \\ 
      HD 72905&   0.04&$\pm$0.07 &   0.02&$\pm$0.06 &  -0.02&$\pm$0.08 &  -0.04&$\pm$0.09 &  -0.18&          &  -0.10&$\pm$0.00 &  -0.14&$\pm$0.04 \\ 
      HD 84737&   0.14&$\pm$0.10 &   0.33&$\pm$0.08 &   0.14&$\pm$0.10 &   0.05&$\pm$0.04 &   0.06&$\pm$0.01 &   0.14&$\pm$0.03 &   0.13&$\pm$0.05 \\ 
     HD 109358&  -0.04&$\pm$0.11 &  -0.08&$\pm$0.05 &  -0.12&$\pm$0.09 &  -0.20&$\pm$0.08 &  -0.20&          &  -0.22&$\pm$0.01 &  -0.22&$\pm$0.04 \\ 
     HD 110897&  -0.49&$\pm$0.10 &  -0.46&$\pm$0.11 &  -0.52&$\pm$0.10 &  -0.51&$\pm$0.05 &  -0.92&          &  -0.62&$\pm$0.04 &  -0.62&$\pm$0.09 \\ 
     HD 137510&   0.28&$\pm$0.09 &   0.46&$\pm$0.12 &   0.30&$\pm$0.10 &   0.23&$\pm$0.10 &   0.34&          &   0.29&$\pm$0.04 &   0.35&$\pm$0.05 \\ 
     HD 141004&   0.10&$\pm$0.11 &   0.19&$\pm$0.12 &   0.05&$\pm$0.06 &  -0.06&$\pm$0.05 &   0.00&          &  -0.00&$\pm$0.05 &   0.01&$\pm$0.03 \\*
     HD 161797&   0.37&$\pm$0.07 &   0.58&$\pm$0.14 &   0.44&$\pm$0.08 &   0.27&$\pm$0.05 &   0.26&$\pm$0.08 &   0.45&$\pm$0.05 &   0.35&$\pm$0.05 \\*
     HD 188512&  -0.04&$\pm$0.05 &   0.05&$\pm$0.12 &   0.00&$\pm$0.09 &  -0.15&$\pm$0.09 &  -0.20&          &   0.01&$\pm$0.07 &  -0.18&$\pm$0.04 \\*

\enddata
\end{deluxetable}